%% file: main.tex
\documentclass[letterpaper,twocolumn,10pt]{article}
\usepackage{usenix}
\usepackage{filecontents}

\usepackage{algorithmic}
\usepackage{textcomp}

\usepackage{times}  
\usepackage{hyperref}
\usepackage{tikz}
\usepackage{amsmath}
\usepackage{amssymb}
\usepackage{graphicx}
\usepackage{xspace}
\usepackage{xcolor}
\usepackage[T1]{fontenc}
\usepackage{subfigure}
\usepackage{verbatim}
\usepackage{url}
\usepackage{enumitem}
\usepackage{listings}
\usepackage{multirow}
\usepackage{multicol}
\usepackage{array}
\usepackage{diagbox}
\usepackage[ruled,linesnumbered]{algorithm2e}
\usepackage{pifont}
\usepackage{mathtools}
\usepackage{fancyhdr}
\usepackage{caption}
\usepackage{xurl}

\newcommand{\crunch}{\vspace{-1mm}}

\newcommand{\sysname}{\texttt{Albireo}\xspace}

\newcommand{\para}[1]{\noindent\textbf{#1}}

\newcommand{\num}[1]{\normalsize{\textcircled{\scriptsize{#1}}}\normalsize\xspace}

\begin{document}

\date{}

\title{Scaling LLM Inference Beyond Amdahl’s Limits via \\Eliminating Non-Scalable Overheads}

\author{
{\rm Alan Zhao}\\
scitix
\and
{\rm Cyril Y. He}\\
scitix
\and
{\rm Wei Xu}\\
Tsinghua University
} 

\maketitle

\begin{abstract}

Deployers of online LLM services usually seek to maximize \emph{cluster-wide} performance given a fixed number of GPUs. Tensor parallelism (TP) is necessary to fit modern models but scales \emph{sub-linearly} as the TP degree \(t\) grows, due to cross-GPU communication and non-scalable runtime work, as predicted by Amdahl's Law. Conversely, increasing \(t\) improves memory efficiency and alleviates \emph{KV}-cache contention and swapping. We identify and validate an empirical optimal TP degree \(t_e\) that balances these effects. We present \sysname, a parallel inference system that \emph{raises the attainable \(t_e\)} by shrinking the non-scalable portion via overlap of scheduling and I/O with compute and sequence-parallel sampling, without changing model architectures. Across models and benchmarks, \sysname achieves up to \(1.9\times\) higher throughput, \(48\%\) lower latency, \(28\%\) higher GPU utilization, and \(54\%\) lower energy than vLLM; in production it yields up to \(2\times\) higher throughput.

\end{abstract}
\maketitle

\input{1_introduction}

\input{2_background}

\input{3_motivation}

\input{4_architecture}
\input{5_async}
\input{6_sampling}
\input{7_evaluation}

\input{8_conclusion}
\bibliographystyle{plain}
\bibliography{ref}

\clearpage

\appendix

\input{Appendix}

\end{document}

%% file: 1_introduction.tex
\section{Introduction}
\label{sec:intro}

In data centers, large transformer-based language models (LLMs)~\cite{ouyang2022gpt, liu2024deepseekv3, vllm_llama3, qwen2.5} require substantial GPU resources to deliver high-throughput, low-latency online inference. 
Deployments are typically constrained by a fixed GPU budget, and operators aim to maximize cluster-wide efficiency—high throughput subject to a latency SLO—for a given number of GPUs. A natural strategy is to run one \emph{instance} per GPU, avoiding cross-GPU collectives and simplifying scheduling. This is usually infeasible because the combined memory footprint of the model weights and the runtime \emph{key–value} (KV) cache~\cite{vaswani2017transformer, prabhu2025vattention, kwon2023vllm} exceeds a single GPU’s capacity. Weight memory scales roughly linearly with model size—for example, a 32B-parameter model in FP16 uses about 64 GB for weights alone—while the KV cache grows with sequence length and batch size and can exceed the weight footprint~\cite{kwon2023vllm, zheng2024sglang, qin2025mooncake}. Given that even high-end GPUs (e.g., H100) provide at most 80 GB per device, fitting 32B or larger models on a single GPU is challenging. To mitigate these constraints, serving systems commonly employ \emph{tensor parallelism} (TP)\footnote{Pipeline parallelism (PP) is suboptimal and reserved for ultra-large models, such as the 1TB Kimi K2~\cite{team2025kimik2}, which span multiple nodes. This paper focuses on intra-node parts of inference systems, so PP is beyond the scope.} to distribute both weights and the KV cache across $t$ (TP degree) GPUs~\cite{vllm_tp, sglang_doc}.

\begin{figure}[tb]
    \centering
    \setlength{\abovecaptionskip}{-5pt}
    \setlength{\belowcaptionskip}{-10pt}
    \includegraphics[width=\columnwidth]{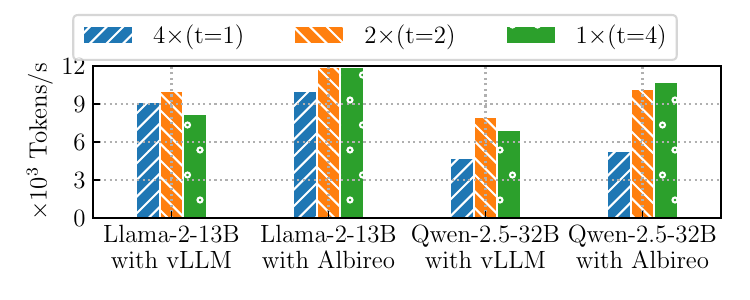}
    \caption{Throughput of different LLMs under various TP degrees ($t$), evaluated with vLLM and \sysname on the $H100^{N}$ testbed (detailed in \S\ref{sec:exp_setup}). Each label $x\times(t{=}y)$ denotes total throughput of $x$ inference instances, each with $t{=}y$. Batch size 128, evenly split (e.g., 64 per instance if 2 instances).}
    \label{fig:tp_effect}
\end{figure}

The main challenge of applying TP is that cross-GPU collectives and \emph{non-parallelizable portion} preclude linear scaling: as the TP degree \(t\) grows, communication increasingly dominates and speedups become sub-linear, with eventual throughput degradation~\cite{shoeybi2019megatron, korthikanti2023megatron2, chitty2024llm}. However, LLM inference is memory-bound; increasing \(t\) expands an instance’s \emph{effective memory budget}, especially for the KV cache, which mitigates request preemption and GPU--CPU KV swapping~\cite{kwon2023vllm, gao2024cost, liu2024cachegen}. The tension between these effects yields an empirical optimum \(t_e\) (per model/workload/hardware). We denote $T(x)$ as the throughput when an instance is deployed with TP degree $x$ and observe \emph{superlinear scaling} from extensive experiments (\S\ref{sec:evaluation}) when $t \le t_{e}$: $T(t) \ge 2\times T(\frac{t}{2})$ and $n\times T(t_{e}) \ge T(n\times t_{e}), n \ge 1$. This indicates, for a fixed GPU budget, configuring each instance with \(t=t_e\) maximizes per-GPU inference performance and thus cluster throughput. Concretely, in Figure~\ref{fig:tp_effect} on a 4-GPU system running vLLM~\cite{kwon2023vllm}, we observe \(t_e=2\); two instances at \(t=2\) outperform four instances at \(t=1\) and a single instance at \(t=4\). Equivalently, total throughput increases with \(t\) for \(t<t_e\) (memory wins) and decreases for \(t>t_e\) (communication wins).
Consequently, for an $N$-GPU cluster, we have
\begin{equation}
\setlength{\abovedisplayskip}{4pt}
\setlength{\belowdisplayskip}{4pt}
\setlength{\abovedisplayshortskip}{2pt}
\setlength{\belowdisplayshortskip}{2pt}
    \label{superlinear}
    \frac{N}{t_{e}} \times T(t_{e}) \geq \frac{N}{t} \times T(t), \quad \forall\, t \in [1, N].
\end{equation}
This implies two paths to maximizing the overall performance: (1) increasing $T(t)$ for a fixed $t$, and (2) increasing $t_{e}$.

However, both directions are challenging. First, increasing $t_e$ is difficult because existing TP implementations~\cite{zheng2024sglang, DeepSpeed, kwon2023vllm} parallelize only the \emph{forward} pass when executing LLMs. According to Amdahl’s Law~\cite{amdahl1967Amdahl} (Acceleration $A = \frac{1}{(1-P) + P/t}$, where $P$ is the \emph{scalable} portion, i.e., \emph{forward}), the overall speedup is bounded by the \emph{non-scalable} fraction $(1-P)$, which becomes more pronounced as $t$ increases. The problem worsens for larger LLMs such as Llama-3.1-70B, where TP is typically set to $t=4$~\cite{vllm_llama3} or $t=8$~\cite{llama2deploy}. Our measurements (\S\ref{sec:evaluation}) show that, under vLLM, throughput at $t=8$ is 21\% lower than $2\times$ that at $t=4$.

Second, increasing $T(t)$ for a fixed $t$ is also non-trivial, especially when $t$ is already near $t_e$. 
Most prior work on inference acceleration tries to increase $T(t)$ by optimizing the \emph{forward} pass. Examples include reducing communication overhead~\cite{kwon2023vllm, zhong2024distserve}, employing KV cache sparsity\cite{bhaskar2025cache, cai2024pyramidkv, li2024snapkv, xiao2023sink, zhang2023h2o} or pre-fetching~\cite{2024InfiniGen,gao2024cost} to minimize cache swapping during forward, and utilizing multi-head latent attention~\cite{liu2024deepseek} for faster forward computations.
However, the \emph{non-scalable} portion $(1-P)$ dominates execution time, so further optimizing the \emph{forward} pass (increasing $P$) via prior techniques or faster GPUs yields diminishing returns (validated in \S\ref{sec:end2end}).

To overcome the speedup limits imposed by Amdahl’s Law, one must reduce the non-scalable portion by increasing the scalable portion $P$ toward 100\%. This paper introduces \sysname, which tackles inference scaling bottlenecks through \emph{overlap} (\S\ref{sec:async_scheduling}, \S\ref{sec:async_inout}) and \emph{parallelization} (\S\ref{sec:sampling}) strategies, enabling faster inference and better resource utilization. Specifically, \sysname overlaps non-scalable scheduling and I/O with computation, and parallelizes sampling across GPUs—both largely under-optimized in prior systems—thereby improving scalability without changing model architectures.

With \sysname, $t_{e}$ shifts to a larger TP degree than that of other engines, and extensive experiments show that it can be estimated using a simple memory-budget rule of thumb (validated in \S\ref{sec:evaluation}):
\begin{equation}
\setlength{\abovedisplayskip}{4pt}
\setlength{\belowdisplayskip}{4pt}
\setlength{\abovedisplayshortskip}{2pt}
\setlength{\belowdisplayshortskip}{2pt}
    \label{tp_selection}
    \begin{aligned}
    t_{e} = \left\lceil 4 \mathcal{M} / \mathcal{C} \right\rceil
    \end{aligned}
\end{equation}
where $\mathcal{C}$ is the per-GPU memory capacity and $\mathcal{M}$ is the LLM weight size. 
As shown in Figure~\ref{fig:tp_effect} (more example in \S\ref{sec:end2end}), \sysname consistently outperforms vLLM for fixed $t$, and can sustain a higher $t_{e}$—e.g., for Qwen-2.5-32B, $t_{e}$ increases from $2$ to $4$. Since throughput exhibits superlinear scaling when $t \le t_{e}$, i.e., $T(t) \ge 2\times T(\frac{t}{2})$, this larger $t_{e}$ directly translates into higher aggregate throughput under a fixed GPU budget.

Our evaluation of \sysname shows substantial gains: $2\times$ higher inference throughput, 48\% lower latency, 28\% higher GPU utilization, and 54\% lower energy consumption compared to vLLM~\cite{kwon2023vllm}, the state-of-the-art inference engine.

In summary, our contributions are:
(1) Performance analysis: We identify key scalability bottlenecks in LLM inference through detailed profiling.
(2) Scalable parallelization: We propose resource-efficient techniques to boost inference scalability and GPU utilization.
(3) System implementation: We develop \sysname, an LLM inference engine featuring asynchronous execution and parallel sampling. 
(4) Comprehensive evaluation: We evaluate \sysname on realistic workloads, demonstrating clear advantages over state-of-the-art systems such as vLLM~\cite{kwon2023vllm} and SGLang~\cite{zheng2024sglang}.

%% file: 2_background.tex
\section{Background}
\subsection{LLM Inference Basics}
\begin{figure}[t]
    \centering
    \setlength{\abovecaptionskip}{-1pt}
    \setlength{\belowcaptionskip}{-10pt}
    \includegraphics[width=\columnwidth]{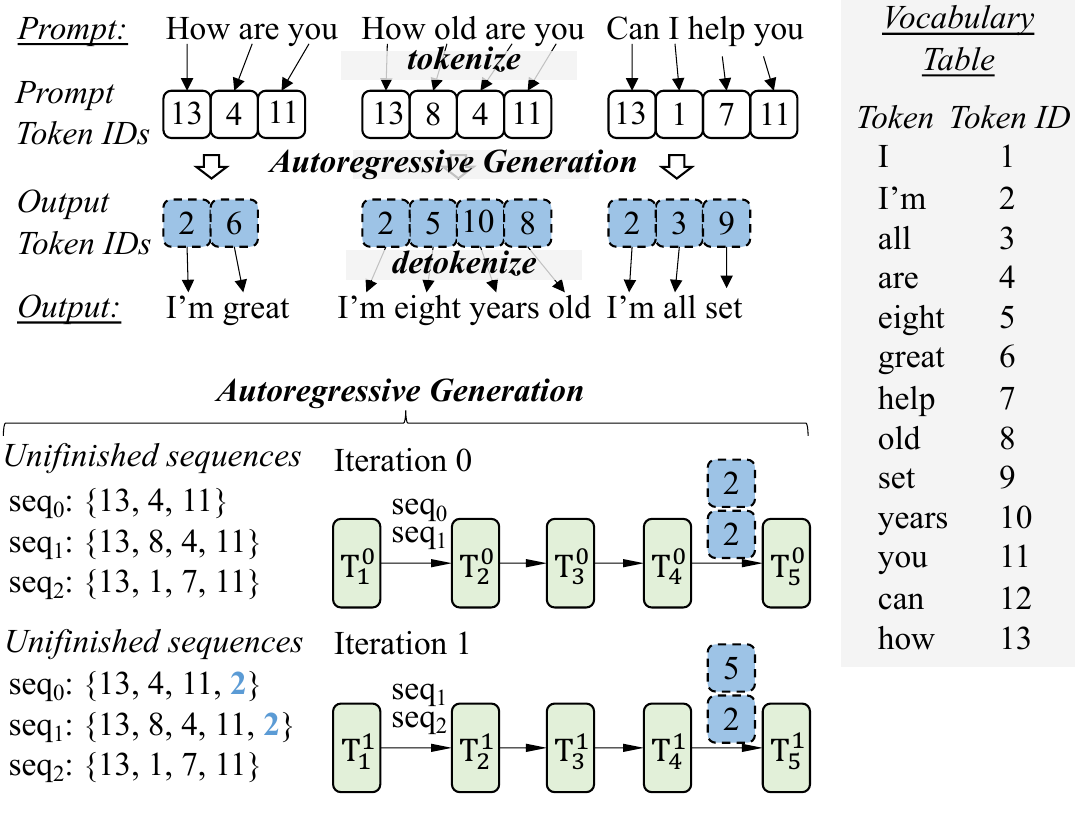}
    \caption{Illustration of the autoregressive generation process.}
    \label{fig:autoregressive}
\end{figure}

\para{\underline{(A) LLM Inference Workflow: }}
\label{sec:background}
\label{sec:back_workflow}
Figure~\ref{fig:autoregressive} overviews autoregressive generation. In LLM inference, an inference engine (e.g., vLLM~\cite{kwon2023vllm}) invokes a trained model to produce an \emph{output} from a user \emph{prompt}. The prompt is tokenized into integer \emph{token IDs} using standard tokenizers~\cite{gage1994new, sennrich-etal-2016-neural, kudo-richardson-2018-sentencepiece}. We call the concatenation of prompt and generated tokens the \emph{sequence}. Inference proceeds through two stages—\emph{prefill} and \emph{decode}—to produce the final output.

\noindent
\emph{\textbf{Prefill.}} In this stage, the model processes the prompt token IDs (e.g., $seq_0$ $\{13,4,11\}$ in iteration 0) in batches through one or more forward passes. The output is a probability distribution over possible next token IDs for each input position.

\noindent
\textbf{Decode.} Starting from the end of the prompt, the model iteratively generates one new token ID at a time. At each step, it uses the prompt and previously generated IDs (e.g., $seq_0$ $\{13, 4, 11, 2\}$ in iteration 1) to compute a probability distribution, from which the next token ID (referred to as the \emph{sampled token ID}) is selected using a sampling strategy (e.g., greedy decoding). This ID is then appended to the sequence. The process repeats until a stopping condition is reached.

\noindent 
\emph{\textbf{Iteration.}} In LLM inference, an \emph{iteration} refers to a cycle of token generation during either the prefill or decode stage. Regardless of the stage, an iteration in popular LLM inference frameworks, such as vLLM~\cite{kwon2023vllm} and SGLang~\cite{zheng2024sglang}, includes five \emph{tasks}. Figure~\ref{fig:autoregressive} also illustrates these tasks. $T_{i}^{n}$ denotes the execution of task $T_{i}$ in the $n$-th iteration.

\noindent
\emph{\textbf{(1) Scheduling ($T_{1}$):}} Existing inference frameworks~\cite{2024Sarathi, yu2022orca, kwon2023vllm} use \emph{iteration batching} for scheduling. At the start of each iteration, the scheduler selects a subset of unfinished sequences based on a priority strategy (e.g., $seq_0$ and $seq_1$ in iteration 0), allocates KV cache blocks for them, and generates \emph{scheduling outputs}. At the end of the iteration, completed sequences are removed, and new ones are added, reducing waiting times for pending sequences.

\noindent
\emph{\textbf{(2) Input processing ($T_{2}$):}}
This task computes the metadata required for model execution and converts the scheduling outputs into \emph{model input}, such as updating cache, transforming lists into tensors and transferring tensors to the GPU. When using TP, it also broadcasts the computed results to all GPUs.

\noindent
\emph{\textbf{(3) Forward ($T_{3}$):}}
Once the model input is prepared, the model computes a probability distribution over possible output tokens, which is then passed to the next task as \emph{logits}. 

\noindent
\emph{\textbf{(4) Sampling ($T_{4}$):}}
Sampling selects the next token ID based on the predicted probability distribution over the \emph{vocabulary}, which is the set of tokens the model can understand and generate. The selected ID(s) are chosen using strategies like greedy search, top-$k$, or nucleus sampling, balancing accuracy and diversity. For example, $\{2\}$ for $seq_0$ in Iteration 0.

\noindent
\emph{\textbf{(5) Output processing ($T_{5}$):}}
This task appends the sampled token ID to the sequence and converts it into a human-readable string through \emph{detokenization}. It also checks for termination conditions and updates the sequence state.

\para{\underline{(B) Key-value Cache:}}
\label{key-value-cache}
In LLM inference, the key-value (KV) cache stores runtime states that are reused in computation, improving throughput and reducing latency. For example, in iteration 1 of $seq_{1}$ in Figure~\ref{fig:autoregressive}, the KV cache prevents the engine from recomputing intermediate results for ${13, 8, 4, 11, 2}$. However, the KV cache is memory-intensive (e.g., about 2.5MB per token ID in Qwen-2.5-32B). Given sequences with thousands of tokens~\cite{databricks} and large batch sizes (e.g., 256), the total GPU memory required can quickly reach hundreds of GB. This makes scaling GPU resources essential for efficient inference.

\para{\underline{(C) Tensor Parallelism in LLM Inference:}}
\label{parallel_inference_tech}
Tensor parallelism (TP)~\cite{shoeybi2019megatron} is a widely used technique to expand KV cache capacity in LLM inference~\cite{vllm_tp, sglang_doc}. It distributes model weights across multiple GPUs, with each GPU computing for only a portion of the model weights. This eliminates redundant weight loading and improves memory efficiency, enabling the model to support more concurrent requests without exceeding the memory capacity. In contrast, data parallelism, which is functionally equivalent to TP when $t=1$, replicates the full model across GPUs and partitions the input data.
For example, with two GPUs each equipped with 80\,GB of memory, a TP degree of \( t = 1 \) leaves \( 2 \times (80 - 64) = 32 \)\,GB available for KV cache when serving Qwen-2.5-32B. In contrast, with \( t = 2 \), the model weights are split across GPUs, leaving \( 2 \times 80 - 64 = 96 \)\,GB available for KV cache.

\subsection{Related Work}
Researchers expand LLM inference parallelism by attacking serialization in kernels and schedulers: Llumnix and SGLang pursue asynchronous execution to curb GPU idle time~\cite{sun2024llumnix, zheng2024sglang}. Kernel/runtime advances push this further—POD-Attention fuses prefill and decode to enable full overlap~\cite{Kamath2025PODAttention}, vAttention uses CUDA virtual memory for dynamic attention-state management~\cite{prabhu2025vattention}, and FlashInfer supplies customizable high-performance attention/sampling kernels~\cite{Ye2025FlashInfer}. Complementary serving work improves throughput–latency trade-offs and long-context scaling: DistServe and Sarathi-Serve separate prefill/decoding~\cite{zhong2024distserve, 2024Sarathi}, ServerlessLLM trims checkpoint loading latency~\cite{fu2024serverlessllm}, and $\mu$serve partitions Transformer modules across devices~\cite{2024userve}. KV-cache methods~\cite{cai2024pyramidkv, li2024snapkv, xiao2023sink, zhang2023h2o} compress or share state, RadixAttention in SGLang~\cite{zheng2024sglang}, and multi-tier designs InfiniGen and CachedAttention~\cite{2024InfiniGen, gao2024cost}. Newer techniques~\cite{Jo2025FastKV, Yuzuguler2025PRESERVE, EntezariZarch2025DEL, Xu2025SpecEE} further shrink or hide movement/compute by compression, prefetch, and speculative early-exit. As a supplement, \sysname targets dependency elimination across I/O, prefill, decode, and sampling to make all tasks parallelizable or overlappable.

Within this line of work, non-scalable overheads have also drawn increasing attention. Recently, several systems reduce CPU–GPU synchronization costs through engineering-level asynchrony—e.g., SGLang’s zero-overhead scheduler~\cite{sglang_zero}, NanoFlow’s asynchronous pipelines~\cite{zhu2025nanoflow}, and vLLM’s emerging asynchronous scheduler~\cite{vllm_async}. Although these efforts reduce certain partial idle-time bottlenecks in $T_{1}$, they do not achieve full asynchrony across $T_{2}$/$T_{4}$/$T_{5}$ due to the absence of task-decoupling mechanisms, nor do they introduce new opportunities for parallel execution. Consequently, their performance gains are limited ($1.8\sim 7.1$\%~\cite{vllm_async}). In contrast, \sysname eliminates all major sources of serialization and enables principled parallelism across tasks ($T_{1}/T_{2}/T_{4}/T_{5}$).

%% file: 3_motivation.tex

\section{Design Overview}
\subsection{Motivation, Challenges and Solutions}
\label{sec:motivation}
\label{sec:challenges} 


\begin{figure}[t]
    \centering
    \setlength{\abovecaptionskip}{-1pt}
    \setlength{\belowcaptionskip}{-10pt}
    \includegraphics[width=0.9\columnwidth]{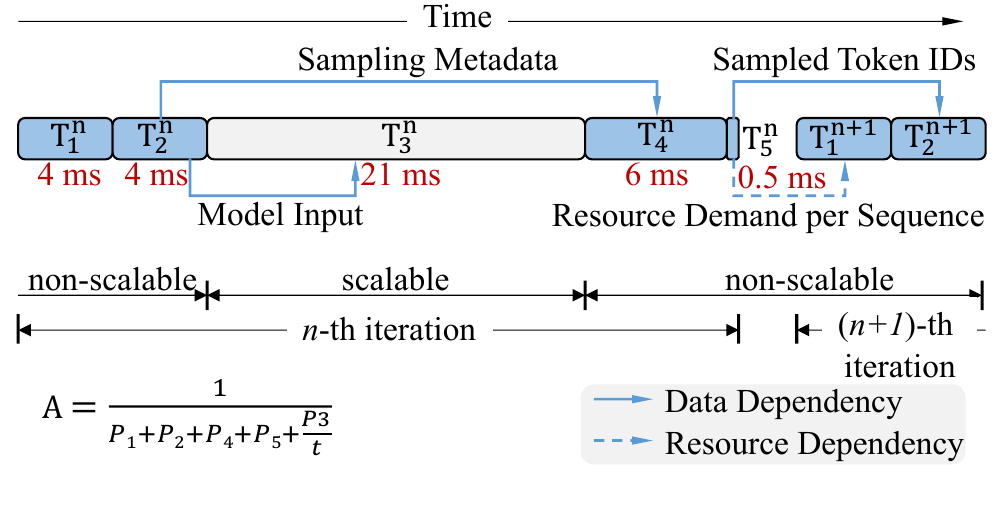}
    \caption{Sequential workflow of inference. Only $T_3^n$ is parallelizable due to dependency constraints. The timing shown is per iteration, and $P_i$ represents the average time proportion of $T_i$. Measurements are based on vLLM 0.11.2 ($t=4$, batch size = 128) running Qwen-2.5-32B on the $H100^{N}$ testbed.}
    \label{fig:vllm_workflow}

\end{figure}

Taking vLLM as an example, we investigate why it fails to achieve a larger $t_{e}$, as estimated by Equation~\ref{tp_selection}; specifically, we profile Qwen-2.5-32B inference with $t=4$.
Figure~\ref{fig:vllm_workflow} shows non-scalable tasks account for $\frac{4 + 4 + 6 + 0.5}{4 + 4 + 21 + 6 + 0.5} \approx 41\% $ of each iteration, forming a throughput bottleneck—a trend consistent across models (\S\ref{sec:end2end}).
According to Amdahl's law, the inference speedup is bounded by \(\frac{1}{P_{1} + P_{2} + P_{4} + P_{5} + P_{3}/t} \). On modern GPUs like the H100, the non-scalable tasks dominate execution time, so further \( P_3 / t \) shrinking (via larger t or faster GPUs) yields little gain.
\sysname addresses this by minimizing non-scalable overhead. It classifies such overhead as either \emph{overlappable} (CPU $T_{1}$, $T_{2}$, $T_{5}$) or \emph{non-overlappable} ($T_{4}$), overlapping the former with $T_{3}$ and introducing a parallel mechanism for $T_{4}$. These efforts yield three key optimizations that tackle the core inference bottlenecks.

\para{Challenge 1: \emph{Resource dependencies in scheduling.}} To address the challenges from non-scalable overhead, we formalize scheduling and make its resource dependencies explicit (dashed arrows in Figure~\ref{fig:vllm_workflow}). A scheduler must respect three system \emph{budgets} during inference: the per-iteration new tokens \( B_{t} \), the number of concurrent sequences \( B_{seq} \), and GPU memory.
With PagedAttention~\cite{kwon2023vllm}, the KV cache is grouped into OS-like \emph{blocks} of every \( B_{c} \) tokens. Therefore, the count of free blocks \( B_{b} \) serves as the effective memory budget and directly shapes scheduling.
Let $\mathbb S$ be current sequences; each $seq$ has length $L_{seq}$ and needs $N_{seq}$ new tokens. The framework chooses $\mathbb S'\subseteq\mathbb S$ to maximize its size subject to
\begin{equation}\label{sche_condition}
\setlength{\abovedisplayskip}{4pt}
\setlength{\belowdisplayskip}{4pt}
\setlength{\abovedisplayshortskip}{2pt}
\setlength{\belowdisplayshortskip}{2pt}
\begin{aligned}
& \max\ |\mathbb S'| \ \text{s.t.}\
\mathbb S'\subseteq \mathbb S ,\
|\mathbb S'|\le B_{seq},\\
 \sum_{seq\in\mathbb S'} & N_{seq}\le B_t ,\
\sum_{seq\in\mathbb S'} \left\lceil \frac{L_{seq}+N_{seq}}{B_c} \right\rceil \le B_b.
\end{aligned}
\end{equation}

Asynchronous scheduling complicates deriving the correct set $\mathbb{S}$ because the previous iteration may still be in flight. Some sequences in $\mathbb{S}$ could finish during output processing of the previous iteration and change state. Since $\mathbb{S}$ affects the scheduler's selection of $\mathbb{S^{'}}$, stale membership can violate the resource constraints in Equation~\ref{sche_condition}. Additionally, misestimating $L_{seq}$ and $N_{seq}$ in asynchronous scheduling can likewise risks breaching these resource budgets.


\para{Optimization 1: \emph{\sysname introduces optimistic single-iteration prediction to enable asynchronous scheduling.}} 
\sysname adopts an \emph{optimistic prediction} strategy (\S\ref{sec:async_scheduling}) to break the resource dependencies between adjacent tasks. In a nutshell, it allocates memory under the assumption that requests never encounter stop conditions and computes $L_{seq}$ and $N_{seq}$ based on this assumption. This enables the scheduling of the next iteration before the current one completes.

\para{Challenge 2: \emph{Data dependencies in input/output processing.}} 
Cross-task data dependencies (solid arrows in Figure~\ref{fig:vllm_workflow}) serialize execution: the input processing of the $(n+1)$-th iteration depends on the output of the $n$-th iteration's output processing to generate the correct model input for \emph{forward}, while that output processing, in turn, waits on non-overlappable sampling to produce token IDs. These dependencies prevent overlapping input/output processing with compute, thereby limiting parallelism and reducing overall efficiency.

\para{Optimization 2: \emph{\sysname introduces early-feedback backfill to enable asynchronous input and output processing.}} 
\emph{Early-feedback backfill} (\S\ref{sec:async_inout}) establishes a fast path from sampling to input and output processing, enabling sampled token IDs to be forwarded immediately upon generation. This is based on the observation that data dependencies are minimal—typically just sampled token IDs. As a result, \sysname can initiate output processing without waiting for sampling to finish, and input processing can proceed concurrently. This overlapping of tasks improves overall inference efficiency.

\para{Challenge 3: \emph{Scaling communication on data dependencies of sampling.}} Original sampling process can be formulated as:
\begin{equation}
\setlength{\abovedisplayskip}{4pt}
\setlength{\belowdisplayskip}{4pt}
\setlength{\abovedisplayshortskip}{2pt}
\setlength{\belowdisplayshortskip}{2pt}
\label{ori_sample}
\begin{aligned}
    [Pr_1^v, Pr_2^v] & = DecodingForward(X) \\
    Pr & = g(Pr_1^v, Pr_2^v) \\
    Y & = Sample(Pr),
\end{aligned}
\end{equation}
where $X$ is the model input, $Pr$ is the logits matrix representing the probabilities of $v$ vocabulary tokens for $s$ requests from \emph{forward}, $g$ signifies the \emph{gather} operation, and $Y$ is the sampling results. The superscript $({\cdot})^v$ denotes partitioning by vocabulary. Sampling remains single-threaded in existing frameworks for two reasons. First, batch-parallel sampling ideally parallelizes along the request (batch) dimension, but the computation is over the vocabulary dimension, which TP already shards. Involving all TP workers requires expensive collectives, adding millisecond-scale overhead each iteration. Second, sampling requires per-sequence metadata (solid arrows in Figure~\ref{fig:vllm_workflow}), such as \emph{penalties}~\cite{penalty-fre-pre, penalty-rep}. In our measurement, serializing and broadcasting this data—e.g., via Pickle~\cite{pickle}—costs more than 10 ms per iteration and causing up to 37\% throughput loss on Qwen-2.5-32B in vLLM.

\para{Optimization 3: \emph{\sysname introduces sequence-parallel sampling.}}
We propose \emph{sequence-parallel sampling} (\S\ref{sec:sampling}), which partitions sampling along the batch dimension so workers can sample independently. It is fully TP-compatible, requiring no model reloads or extra GPU memory.
By \emph{overlapping scatter} of sampling metadata and exchanging logits via \emph{all-to-all}, \sysname removes the communication bottleneck that limits parallel sampling efficiency.

%% file: 4_architecture.tex

\subsection{Overview}
\label{sec:architecture}
\begin{figure}[t]
    \centering
    \setlength{\abovecaptionskip}{-1pt}
    \setlength{\belowcaptionskip}{-10pt}
    \includegraphics[width=0.9\columnwidth]{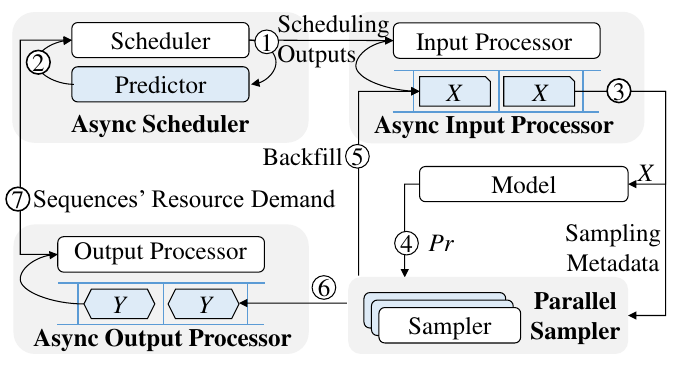}
    \caption{Architecture of \sysname. Hollow components represent the baseline workflow described in \S\ref{sec:back_workflow}; solid components enable asynchronous and parallel execution.} 
    \label{fig:archi}

\end{figure}

\textbf{Architecture}: \sysname makes minimal changes to existing inference stacks by adding four components: an \emph{asynchronous scheduler}, \emph{input processor}, \emph{output processor}, and \emph{parallel sampler} (Figure~\ref{fig:archi}). The \emph{asynchronous scheduler} incorporates a \emph{predictor} that estimates resource needs from current scheduling outputs, allowing the next round to start before the current iteration finishes. The \emph{input processor} asynchronously queues model inputs ($X$) and consumes them in order. The \emph{parallel sampler} launches multiple sampler instances to partition the sampling task; their results ($Y$) are enqueued to the ordered queue of \emph{output processor} for detokenization.

\noindent
\textbf{Workflow:}
In this section, we illustrate the workflow of \sysname through an example in which the $n$-th iteration of \emph{forward} commences on the GPU while the scheduling for the $(n+1)$-th iteration simultaneously initiates on the CPU. 

On the CPU, the scheduler inspects sequence states and, under a policy, e.g., First-Come-First-Serve (FCFS), emits scheduling outputs for $n{+}1$, forwarding them to the predictor and input processor (\num{1}). The predictor estimates the KV blocks needed in the next iteration per scheduled sequence and pre-updates their states (\num{2}). In parallel, the input processor converts these outputs into model inputs and sampling metadata in a queue.

On the GPU, \sysname dequeues iteration-$n$ inputs and sampling metadata and dispatches them concurrently to the model and the parallel sampler (\num{3}). After \emph{forward}, the sampler shards the model \emph{logits} $Pr$ to sampler instances (\num{4}), which perform parallel sampling using the metadata.

As each token ID is produced, the sampler \emph{backfills} it to the input processor to complete iteration-$n{+}1$ inputs (\num{5}), resolving placeholders inserted when a sequence spans iterations and its last token was unknown at scheduling. When sampling finishes, the sampler sends results to the output processor (\num{6}).
Simultaneously, the output processor continuously extracts sampling results from the queue, updates the states of unfinished sequences, and removes completed ones (\num{7}).

\para{Optimized inference pipeline.} \sysname’s pipeline is primarily dominated by the model \emph{forward}. Figure~\ref{fig:pipeline} shows the optimized design: relative to the sequential workflow (Figure~\ref{fig:vllm_workflow}), Optimizations 1–2 shrink CPU time from 8.5 ms to $<80 \mu$s, and Optimization 3 cuts sampling on four H100s from 6 ms to 1.5 ms. These remove $>89\%$ of non-scalable overhead, yielding an $\sim1.7\times$ speedup over vLLM.

%% file: 5_async.tex
\begin{figure}[t]
    \centering
    \setlength{\abovecaptionskip}{-1pt}
    \setlength{\belowcaptionskip}{-10pt}
    \includegraphics[width=0.9\columnwidth]{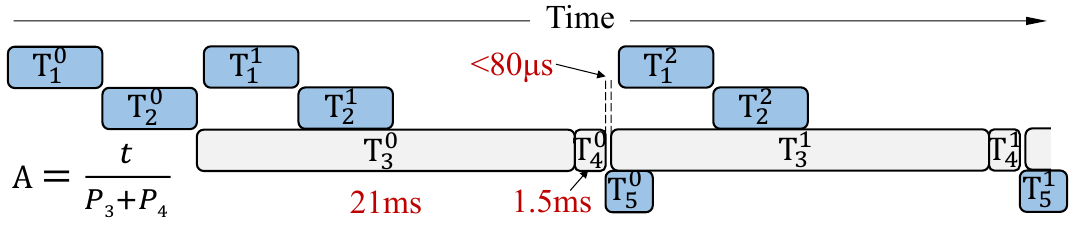}
    \caption{\sysname's execution pipeline. Conducted on Qwen-2.5-32B with 4$\times$80GB H100 (t=4, batch size=128).}
    \label{fig:pipeline}
    \crunch
    \crunch
    \crunch
\end{figure}

\section{Optimistic Asynchronous Scheduling}
\label{sec:async}

\label{sec:async_scheduling}

In \emph{iteration batching} scheduling, the $(n+1)$-th iteration depends on the completion of the $n$-th iteration. This synchronous scheduling enforces strict sequential execution, preventing asynchronous input/output processing. 

Upon re-examining Equation~\ref{sche_condition}, asynchronous scheduling still appears feasible. Before a sequence \(seq\) is finished, we can easily deduce that \(L_{seq}^{n+1} \gets L_{seq}^{n} + N_{seq}^{n}\), where the initial length of \(L_{seq}^{0}\) is 0, and \(N_{seq}^{n}\) can be inferred based on the sequence's stage (prefill or decode), which is determined by \(L_{seq}^{n}\). The remaining challenge lies in predicting whether a sequence is finished in asynchronous scheduling, ensuring the accurate identification of the sequence set $\mathbb{S}$.

\noindent
\textbf{Iteration-dependent sequence management.}
We first present a revised sequence management mechanism tailored for asynchronous scheduling. As Figure~\ref{fig:pipeline} shows, 
during $T_1^{n+1}$, a sequence  \(seq\) involving in the $(n-1)$-th iteration may exhibit two uncertain length: \(L_{seq}^{n-1}\) prior to \(T_5^{n-1}\) and \(L_{seq}^{n-1} + N_{seq}^{n-1}\) after \(T_5^{n-1}\). This dual-state condition creates potential ambiguities for the scheduler. 
Enforcing the sequential execution of \(T_1^{n+1}\) and \(T_5^{n-1}\) in any order undermines the advantages of asynchrony and introduces unnecessary event management overhead, reducing inference performance.

To resolve sequence state inconsistencies, \sysname introduces \emph{iteration-dependent sequence management}, which assigns a virtual state to each token ID in every sequence. 
The asynchronous scheduler tracks three states for each sequence during the $n$-th iteration: (1) \emph{Expected length (EL)}: The sequence length at the start of the $n$-th iteration; (2) \emph{Current length (CL)}: The sequence length at the end of the $n$-th iteration; (3) \emph{Number of new token IDs (NNT)}: The number of tokens expected to be generated by the end of the $n$-th iteration.
This scheme lets the scheduler query future states based on how many iterations a sequence has been scheduled, ensuring consistent state tracking during asynchronous execution of \(T_1^{n+1}\).



\noindent
\textbf{Optimistic prediction.}
The essence of asynchronous scheduling involves two key aspects: (\textbf{A1}) determining the number of KV cache blocks required by a sequence in each scheduling iteration, and (\textbf{A2}) predicting whether the sequence should continue generating the next token ID (i.e., whether it needs further processing).
The core of \textbf{A1} is to compute the number of KV cache blocks \(C_n\) needed for a sequence, given its scheduled iteration \(n\). \(C_n\) can be calculated interactively as follows:
\begin{equation}\label{cal_block_num}
\setlength{\abovedisplayskip}{4pt}
\setlength{\belowdisplayskip}{4pt}
\setlength{\abovedisplayshortskip}{2pt}
\setlength{\belowdisplayshortskip}{2pt}
C_n=\left\lceil\frac{L_n}{B_c}\right\rceil,\qquad
L_n=\begin{cases}
L_{n-1}+1, & L_{n-1}\ge N_p,\\
L_{n-1}+N_c, & L_{n-1}<N_p.
\end{cases}
\end{equation}
Here, $L_{0}=0$, \(N_p\) represents the number of prompt token IDs in the sequence, and \(N_c\) denotes the chunk size of token IDs processed in prefill. During the prefill stage (when \(L_{n-1} < N_p\)), \sysname updates \(L_n\) based on the preset \(N_c\) to enhance the efficiency of the prefill process. In contrast, during the decode stage, only one token ID is generated per iteration, so it is sufficient to increase the current length by only one.
For \textbf{A2}, \sysname \emph{optimistically} predicts that each sequence requires further inference. This strategy is highly effective in practice because, for each prompt generating \(N\) tokens, the prediction succeeds \(N-1\) times and fails only once.

\sysname uses \emph{single-iteration} asynchronous scheduling, where only the $(n+1)$-th iteration is scheduled during the $n$-th iteration. This approach offers two benefits: first, we can hide the CPU computation by fully overlapping it with the \emph{forward}, only adding a negligible 80 µs overhead to dequeue model input and enqueue sampling outputs (Figure~\ref{fig:pipeline}).
Second, in online serving, when a new request arrives, \sysname must recompute the scheduling for prefill tasks to ensure consistent latency for the generation of the first token. Consequently, the more iterations advanced in asynchronous scheduling, the higher the likelihood of scheduling invalidation. Single-iteration asynchronous scheduling strikes an effective balance between computational overhead and performance gains.



\begin{figure}[t]
    \centering
    \setlength{\belowcaptionskip}{-10pt}
    \includegraphics[width=0.9\columnwidth]{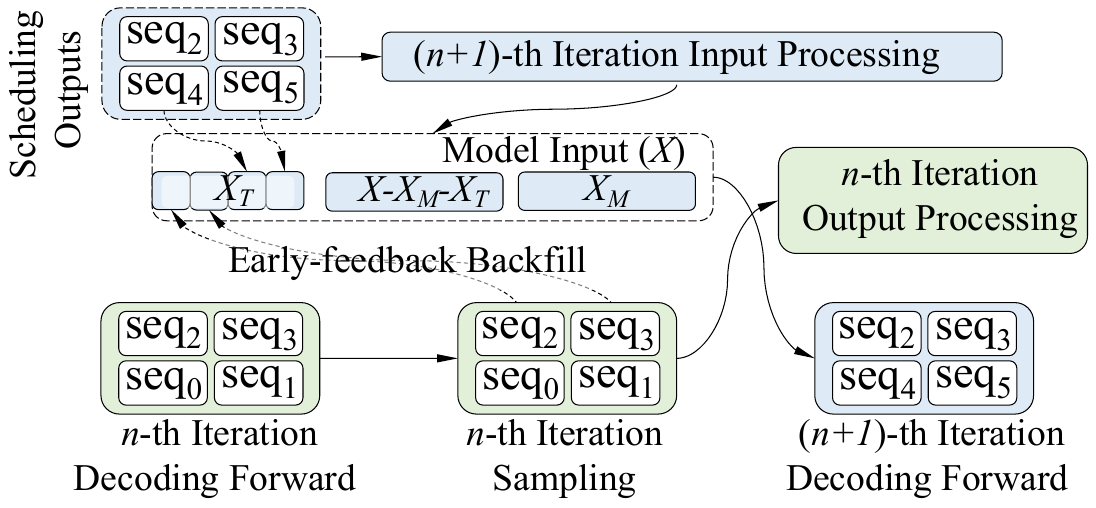}
    \caption{Illustration of asynchronous input processing.}
    \label{fig:input_proc}

\end{figure}

\section{Asynchronous Input/Output Processing}
\label{sec:async_inout}
By decoupling scheduling from the sequential execution workflow as an asynchronous operation, input processing can also be performed asynchronously in advance. However, since the input processor also depends on the outputs from the output processor to generate model input, and the output processor in turn relies on the sampling, challenges remain for achieving asynchronous input/output processing fully.

\begin{figure}[t]
    \centering
    \setlength{\belowcaptionskip}{-10pt}
    \includegraphics[width=0.8\columnwidth]{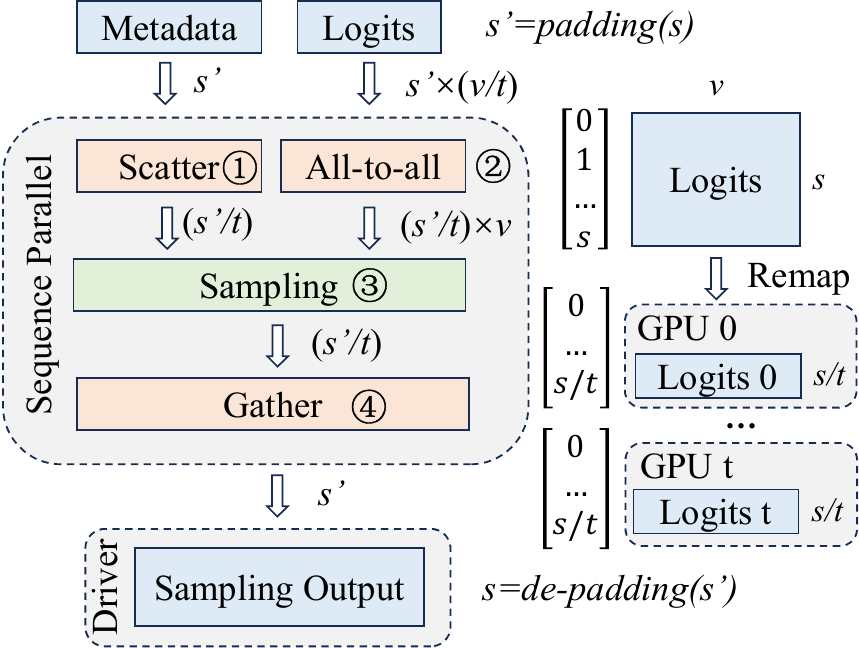}
    \caption{Sequence-parallel sampling workflow.}
    \label{fig:sample}
    \crunch
    \crunch
    \crunch
\end{figure}

\noindent
\textbf{Early-feedback backfill.}
When generating the $(n+1)$-th token ID in the $(n+1)$-th iteration, the model requires the $n$-th token ID of the sequence as part of inputs. 
Without this $n$-th token ID, the model will produce incorrect token IDs, resulting in \emph{inference errors}.

To further decouple input processing and output processing from the sequential execution workflow and eliminate the dependency of input processing on output processing, \sysname introduces an \emph{early-feedback backfill} mechanism. This mechanism establishes a fast path from the sampler to both processors. During sampling, each newly generated token ID is immediately forwarded to these processors, which aligns model inputs with those in the sequential execution workflow and corrects potential inference errors.

\noindent
\textbf{Input processing.}
The input processing of the $(n+1)$-th iteration relies on both the scheduling output of the $(n+1)$-th iteration and the sampling output of the $n$-th iteration. These dependencies are the reason why existing inference frameworks cannot execute input processing asynchronously. In \sysname, the input processor can directly access the output from the asynchronous scheduling, reducing the data dependency to only the sampling output from the $n$-th iteration.

As Figure~\ref{fig:input_proc} shows, a model input $X$ comprises metadata ($X_{M}$) and tensors ($X-X_{M}$). The input processor's dependency on the sampler affects only one tensor in $X-X_{M}$—the \emph{last sampled token IDs} ($X_{T}$). To prevent input processing from stalling due to unavailable sampling output from the previous iteration, \sysname divides input processing into three steps. 
First, it computes $X_{M}$, such as the metadata used for KV cache. 
Second, it allocates every tensor in $X-X_{M}$, including input positions and the last sampled token IDs. 
Third, it transfers $X-X_{M}-X_{T}$ to GPUs.  
Although the content of $X_{T}$ depends on sampling outputs, its shape is determined exclusively by scheduling outputs alone. This allows the input processor to execute the three processing steps as soon as it receives scheduling outputs, without waiting for sampling. \sysname corrects $X_{T}$ in two scenarios. If a sequence was not scheduled in the previous iteration, (e.g., \( seq_{4} \) and \( seq_{5} \) in Figure~\ref{fig:input_proc}), its last token ID is used as the sampled token ID. 
Otherwise, the sampler backfills token IDs for these sequences (e.g., \( seq_{2} \) and \( seq_{3} \)) after the sampling process completes.

\noindent
\textbf{Output processing.}
Although \sysname overlaps GPU and CPU tasks extensively, it still requires rapid output processing because the \emph{forward} time may be insufficient to hide the overhead of all CPU tasks~\footnote{The profiling results in Figure~\ref{fig:vllm_workflow} show only the unhidden portion of output processing, while the full process takes approximately 7 ms.}. 
Consequently, \sysname preserves the existing output processing logic while incorporating several engineering optimizations (detailed in Appendix A of the supplementary material), including utilizing two \emph{lookup tables}, replacing \emph{de-tokenizer} calls with fast table lookups, and parallelizing sequence processing and stop checking, reducing output processing time to under 2 ms and thus effectively hiding CPU time in the execution timeline.

%% file: 6_sampling.tex
\section{Sequence-Parallel Sampling}
\label{sec:sampling}

\subsection{Parallel Sampling Workflow}
As highlighted in \S\ref{sec:challenges}, TP focuses exclusively on the computationally intensive \emph{forward} task, but the sampling task only runs on a single GPU. Consequently, the \emph{driver worker} (the first process among all worker processes) alone handles sampling, leaving the GPUs of other workers idle. We notice that in non-parallel sampling, operations on the $(s \times v)$ logits (Figure~\ref{fig:sample}) are independent along the sequence dimension. This independence allows the sampling workload to be partitioned across multiple GPUs, thereby improving scalability.

In contrast to the original workflow described in Equation~\ref{ori_sample}, the sequence-parallel sampling process can be formulated as
\begin{equation}
\setlength{\abovedisplayskip}{4pt}
\setlength{\belowdisplayskip}{4pt}
\setlength{\abovedisplayshortskip}{2pt}
\setlength{\belowdisplayshortskip}{2pt}
\label{opt_sample}
\begin{aligned}
    [Pr_1^v, Pr_2^v] & = DecodingForward(X) \\
    [Pr_1^s, Pr_2^s] & = \hat{g}(Pr_1^v, Pr_2^v) \\
    Y_1^s &= Sample(Pr_1^s) \ and \ Y_2^s = Sample(Pr_2^s) \\
    Y & = g([Y_1^s, Y_2^s]),
\end{aligned}
\end{equation}
where $g$ and $\hat{g}$ are \emph{gather} and \emph{all-to-all} operations, respectively. 
The \sysname sampling workflow includes four steps: 
(1) The driver worker uses \emph{scatter} to distribute sampling metadata of different sequences to workers. 
(2) Each worker performs an \emph{all-to-all} operation to exchange vocabulary probabilities after \emph{forward}, splitting the data along the sequence dimension. 
(3) The core sampling computation remains unchanged, but each worker processes a smaller subset of requests independently. 
(4) The driver worker collects the sampled token IDs from all workers into an array and sends it via \emph{early-feedback backfill}.

\subsection{Key Optimizations}
The parallel sampling workflow poses two primary challenges: (1) the additional communication overhead, which may degrade performance, and (2) the risk of altering the final output due to parallel computation. In this section, we present the key optimizations designed to address both issues and ensure consistent, high-performance results.

\para{Overlapping scattering with GPU computation.} The \emph{forward} preceding sampling is a GPU-bound operation, while the substantial sampling metadata is only needed after the \emph{forward} computation completes. To leverage this, \sysname delays the scatter operation until the \emph{forward} begins, effectively concealing the communication overhead within the inevitable GPU computation time. We define the \emph{ratio} of sampling metadata scattering time to \emph{forward} pass time as $R_s$. In most cases, $R_s$ is lower than 20\% even on H100 GPUs because the metadata size is small ($\sim$1.5 KB per request for a sequence length of 1000). This full overlap effectively eliminates the communication overhead associated with scattering.

\para{Efficient logit exchange.} \sysname adopts a more efficient approach to exchanging logits. In conventional setups, the driver worker gathers probabilities along the vocabulary dimension from all workers for sampling. However, once the sampling workload is distributed across all workers, each worker must access these probabilities. Therefore, \sysname replaces the traditional \emph{gather} with an \emph{all-to-all} operation, which gathers probabilities along the vocabulary dimension and partitions them according to batch size simultaneously. This approach eliminates the need for workers to individually partition the outputs of an \emph{all-gather} operation. As a result, each worker transfers the same amount of data as in a single \emph{gather} operation, ensuring that parallel sampling does not introduce additional communication overhead at this step.

The driver worker collects results only after the sampling task is completed. Each result consists of a single token ID and sequence ID per sequence. As a result, the communication data remains minimal, even for large batch sizes. \emph{We only observe 200 $\mu$s delay on average for 256 requests.}

\para{Deterministic random number generation.} Independent RNG states make per-worker draws diverge from single-worker outcomes. To retain determinism under a fixed seed, \sysname pre-generates all $k$ random numbers on all $t$ GPUs and lets each worker consume its slice. This approach trades additional GPU memory (approximately 128 KB per request) for lower communication overhead (saving about 1 ms to retrieve them from remote GPUs).

\para{Batch padding.}    
Collectives often require divisibility by the number of peers. \sysname applies \emph{batch padding} to make $s' \ mod \ t = 0$ with synthetic metadata and extra logit rows; the driver then discards padded outputs before early-feedback backfill. Although this introduces some additional communication padding overhead (less than 10 KB for metadata padding), the advantages of parallel sampling far outweigh the costs, leading to an overall speedup in inference.

%% file: 7_evaluation.tex
\section{Implementation}
\label{sec:impl}

The core functionality of \sysname is implemented as a framework-independent plugin, comprising approximately 8,500 lines of C++ code. This modular design allows \sysname to be easily incorporated into various frameworks with minimal modifications. For our evaluation, we integrated the plugin with vLLM, requiring only $\sim$200 lines of code changes.
\begin{figure*}[tb]
    \centering
    \setlength{\abovecaptionskip}{-1pt}
    \setlength{\belowcaptionskip}{-10pt}
    \includegraphics[width=0.95\textwidth]{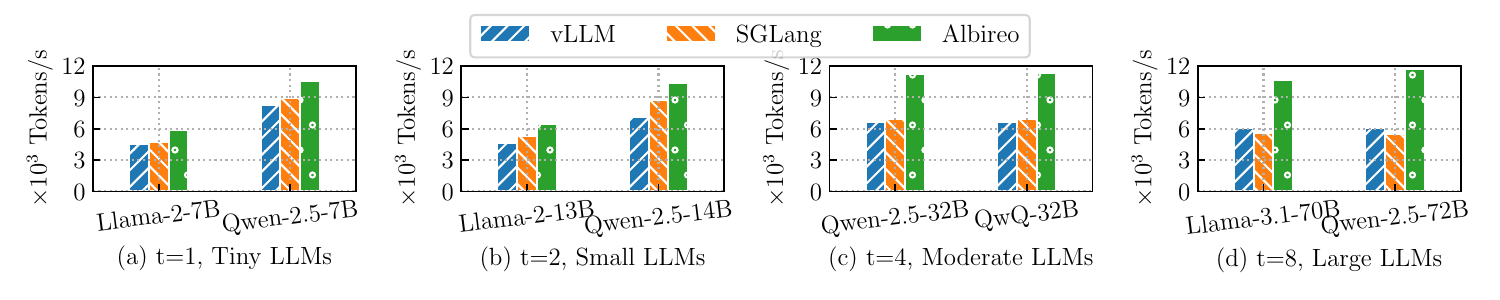}
    \caption{Inference throughput across different model sizes, measured using the default configuration on the $H100^{N}$ testbed.}
    \label{fig:tput_on_model_size}
\end{figure*}

\section{Evaluation}
\label{sec:evaluation}

We evaluate \sysname to answer the following key questions: 
(1) What performance improvements does \sysname deliver, and on which metrics? Is \sysname effective across different models, settings, and GPUs?(\S\ref{sec:end2end}, \S\ref{sec:e2e_latency})
(2) How does \sysname perform in production? (\S\ref{sec:production})
(3) Can \sysname make more efficient use of resources? (\S\ref{sec:resource})
(4) What are the primary sources of performance gain? Does \sysname make the correct design decisions? (\S\ref{sec:studies})


\subsection{Experimental Setup}
\label{sec:exp_setup}
\para{Testbed.}
We evaluate \sysname on three testbeds: $H100^{N}$, $A100^{N}$ and $A100^{P}$. Each has 8 GPUs (80 GB memory): $H100^{N}$ uses NVLink-connected H100 GPUs, $A100^{N}$ uses NVLink-connected A100 GPUs, and $A100^{P}$ uses PCIe-connected A100 GPUs. All testbeds feature 2 TB RAM, an Intel® Xeon® Platinum 8468 CPU with 192 logical cores.

\para{Baseline.} We evaluate \sysname against vLLM (v0.11.2)~\cite{kwon2023vllm} and SGLang (v0.5.5)~\cite{zheng2024sglang}, using their latest stable releases. All frameworks are evaluated with their default configurations, ensuring that all built-in performance optimizations are enabled.
\footnote{Sections~\ref{sec:production} to \ref{sec:studies} compare only against vLLM, since \sysname is implemented as a vLLM plugin, highlighting the incremental impact of its optimizations.}

\para{Models.} We evaluate eight LLMs with FP16 precision spanning a range of commonly used sizes, grouped into four categories: (1) \textit{Tiny} — Llama-2-7B~\cite{llama-2-7b}, Qwen-2.5-7B~\cite{qwen2.5}; (2) \textit{Small} — Llama-2-13B~\cite{llama-2-13b-hf}, Qwen-2.5-14B; (3) \textit{Moderate} — Qwen-2.5-32B~\cite{qwen2.5}, QwQ-32B~\cite{qwq}; and (4) \textit{Large} — Llama-3.1-70B~\cite{llama-3.1}, Qwen-2.5-72B~\cite{qwen2.5}. Larger models such as DeepSeek V3~\cite{liu2024deepseekv3} are not included, as they require PP across multiple nodes, which is beyond our scope.

\para{Configuration.} In all tests, PP is disabled, and DP is expressed as a TP degree of $t=1$. By default, each model category adopts the $t_{e}$ derived from Equation~\ref{tp_selection}, that maximizes overall performance within a node: 1 for \textit{Tiny}, 2 for \textit{Small}, 4 for \textit{Moderate}, and 8 for \textit{Large}. We also vary $t$ in \S\ref{sec:end2end} to validate performance across different TP degrees. Following prior work~\cite{sglang_bench}, we adopt a default per-GPU batch size of 32 (e.g., a total batch size of 128 when $t=4$) to better utilize resources under high-load datacenter settings; \S\ref{sec:production} further presents comparisons under low-load conditions.

\para{Workload.}
We randomly sample prompts from the Databricks dataset~\cite{databricks} as user inputs. For production deployment, we adopt \texttt{bentoML}~\cite{bentoML} to launch servers and clients. All sampling features are enabled, including top-$p$, top-$k$, min-$p$, temperature, and repetition, presence, and frequency penalties.

\para{Metrics.} We evaluate inference performance using four key metrics:  
\emph{Throughput} — the number of tokens generated per second;  
\emph{Latency} — the average time to generate the next token;  
\emph{GPU Utilization} — the efficiency of GPU resource usage;  
\emph{Power Usage} — the real-time GPU power consumption.  



\subsection{End-to-End Throughput Improvement} 
\label{sec:end2end}

\para{Throughput improvements for LLMs of all sizes.} 
We first evaluate the performance of each LLM on the $H100^{N}$ testbed using default configurations, as shown in Figure~\ref{fig:tput_on_model_size}. We observe that \sysname yields improvements of about 1.3$\times$, 1.5$\times$, 1.7$\times$ and 1.9$\times$ for \textit{Tiny}, \textit{Small}, \textit{Moderate}, and \textit{Large} LLMs, respectively. These differences arise from the varying $t_e$ across model sizes. For a given model, \sysname reduces overhead by about $T_{1} + T_{2} + (1 - \frac{1}{t}) \times T_{4} + T_{5}$ in each iteration. For \textit{Tiny} models with $t = 1$, parallel sampling and early feedback in asynchronous input processing ($T_{2}$) are disabled, leading to only partial benefits from asynchronous execution. In contrast, larger models (with $t > 1$) fully exploit these mechanisms because parallel sampling removes roughly $(1 - \frac{1}{t}) \times T_{4}$ of the sampling overhead, leading to greater speedups.  
Models within the same category show similar performance gains, as they use the same TP degree.

\begin{figure}[t]
    \centering
    \setlength{\abovecaptionskip}{-1pt}
    \setlength{\belowcaptionskip}{-10pt}
    \includegraphics[width=0.95\columnwidth]{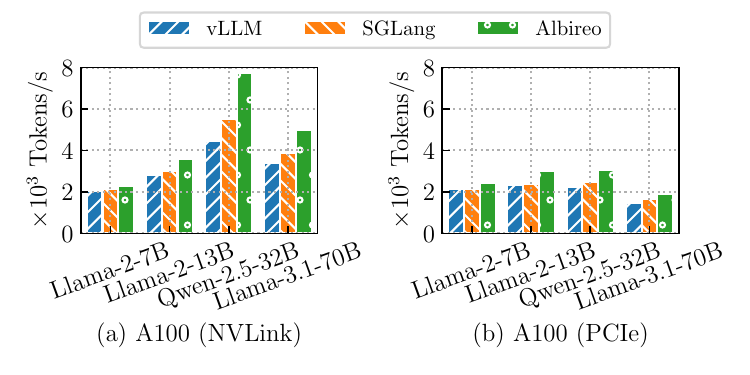}
    \caption{Throughput across different devices.}
    \label{fig:tput_on_device}
    \crunch
    \crunch
    \crunch
\end{figure}
\begin{figure*}[tb]
    \centering
    \setlength{\abovecaptionskip}{-2pt}
    \setlength{\belowcaptionskip}{-10pt}
    \includegraphics[width=0.95\textwidth]{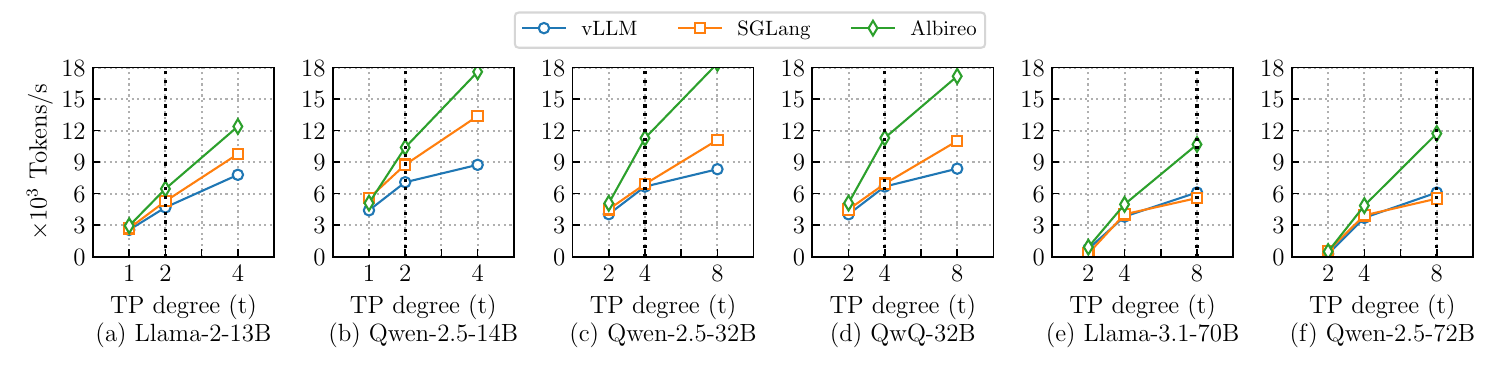}
    \caption{Impact of TP degree on throughput. The dashed line indicates the \( t^{e} \). To the left of the dashed line, \sysname not only outperforms others but also achieves superlinear scalability, with throughput at \( t = 2n \) exceeding $2\times$ that at \( t = n \).}
    \label{fig:tput_superlinear}
\end{figure*}

\noindent
\textbf{Cross-platform validation of \sysname.}
To assess the generality of \sysname, we then evaluate its performance across diverse GPU architectures, conducting experiments on the $A100^{N}$ and $A100^{P}$ testbeds, as shown in Figures~\ref{fig:tput_on_device}(a) and~\ref{fig:tput_on_device}(b). Since models of similar size yield comparable speedups, we report results for one representative model per size category, all measured with the default configuration. \sysname delivers up to 1.7$\times$ and 1.4$\times$ performance speedups over vLLM on $A100^{N}$ and $A100^{P}$, respectively, demonstrating its effectiveness across GPU types.

Speedups are higher on $A100^{N}$ due to faster inter-GPU communication during forward passes when $t > 1$, which reduces $T_{3}$ and improves overall performance.

Compared to H100, A100 shows lower overall gains—an expected result, as H100's 3$\times$ higher FLOPs makes $T_{3}$ less dominant. This trend highlights the increasing impact of non-scalable overhead as computational power grows, reinforcing the importance of \sysname's optimizations.

\noindent
\textbf{Higher aggregate throughput from superlinear scaling and larger $t_{e}$.}
When memory permits, models can be deployed with different TP degrees. For example, on the $H100^{N}$ testbed, a node supports $\frac{8}{t}$ inference engine instances with TP degree $t$. This section investigates whether \sysname can support a larger $t_{e}$ and achieve higher aggregate throughput under the same GPU budget.

Figure~\ref{fig:tput_superlinear} shows the throughput of a single inference engine under different TP degrees, denoted as $E^{t}$. Results for \emph{Moderate} and \emph{Large} LLMs at $t=1$ are omitted: the former exhibits poor performance, while the latter runs out of memory. In both vLLM and SGLang, $E^{t_{e}} < 2 \times E^{t_{e}/2}$, indicating that the actual effective TP degree $t_{e}'$ is lower than $t_{e}$. This trend is more pronounced for \emph{Large} LLMs, where communication and coordination overheads increase with higher TP degrees. Such behavior aligns with Amdahl’s Law, which predicts diminishing returns as parallelism grows due to the impact of communication and other non-scalable tasks.

In contrast, \sysname increases the effective $t_{e}$ from 2 to 4 for \emph{Moderate} LLMs and from 4 to 8 for \emph{Large} LLMs. At the same TP degree $t$, \sysname consistently outperforms the baselines and demonstrates superlinear scaling across all model sizes for $t \le t_{e}$, i.e., $E^{t_{e}} \ge 2 \times E^{t_{e}/2}$, which leads to higher aggregate throughput, i.e., $\frac{8}{t_{e}} \times E^{t_{e}}$. This is primarily because \sysname significantly reduces \emph{non-scalable} bottlenecks and efficiently shares model weights among GPUs at higher TP degrees, thus freeing redundant memory for the KV cache. For memory-bound inference workloads, this enhances cache hit rates and increases resource utilization on GPU compute (validated in \S\ref{sec:resource}), driving superlinear throughput gains. 

\begin{figure}[tb]
    \centering
    \setlength{\abovecaptionskip}{-3pt}
    \setlength{\belowcaptionskip}{-10pt}
    \includegraphics[width=0.95\columnwidth]{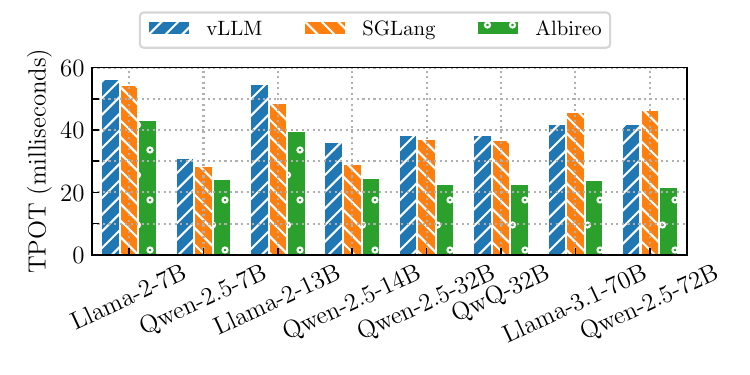}
    \caption{TPOT across different models and engines.}
    \label{fig:tput_tpot}
    \crunch
    \crunch
    \crunch
\end{figure}

\subsection{Reduced End-to-End Latency} 
\label{sec:e2e_latency}

\para{Time-to-First-Token (TTFT).} TTFT is dominated by the prefill phase. In this phase, each batch typically contains thousands of tokens (e.g., 8192), and iterations are largely spent on GPU forward passes; moreover, prefill does not involve sampling. As a result, \sysname's optimizations have minimal effect, and TTFT is nearly identical to vLLM; detailed results are omitted for brevity.

\para{Time-per-Output-Token (TPOT).} TPOT measures the average time to generate each output token ID after the previous one. We evaluate TPOT for each model on the $H100^{N}$ testbed using default configurations, as shown in Figure~\ref{fig:tput_tpot}. \sysname consistently achieves a 22\% to 48\% reduction in average TPOT across different LLMs, with the largest improvement observed on Qwen-2.5-72B.
By reducing non-scalable overhead ($1-P$), \sysname shortens iteration time and lowers token generation latency. Our results show consistent performance gains from \sysname for all scenarios, especially for \textit{Moderate} and \textit{Large} LLMs.

\subsection{Online Inference in Production}
\label{sec:production}
\sysname is integrated into our cloud-based \emph{Model-as-a-Service} (MaaS) platform, enabling users to deploy their online inference tasks like question answering using user-provided models. We evaluate its performance below.

\begin{figure}[t]
    \centering
    \setlength{\abovecaptionskip}{-1pt}
    \includegraphics[width=\columnwidth]{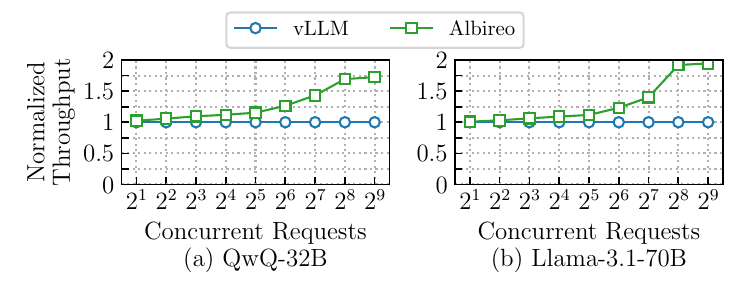}
    \caption{Normalized inference throughput under varying loads on an H100 node, for TP degrees $t{=}4$ (left) and $t{=}8$ (right). Each value is normalized to vLLM under the same load (baseline $=1$).}
    \label{fig:production_tput}
\end{figure}

\para{Throughput under varying load conditions.}
Unlike the controlled benchmarks above, which maintain a large batch size to stress-test inference engines, the load in production environments fluctuates in real time—sometimes underloaded, sometimes overloaded. Figure~\ref{fig:production_tput} illustrates the performance comparison of two popular LLMs on our MaaS platform, which serves users, after transitioning from vLLM to \sysname.

We observe that the performance benefits of \sysname grow with increasing load pressure and eventually stabilize ($1.7\times$ for QwQ-32B and $2\times$ for Llama-3.1-70B). Through deep profiling, we find that before the system becomes overloaded (i.e., concurrent requests $\leq 256$), increasing requests from $n$ to $2n$ results in faster growth of non-scalable tasks ($T_{1}/T_{2}/T_{4}/T_{5}$) than the forward computation ($T_{3}$) because the system is memory-bound. With increasing batch size, non-scalable overheads grow disproportionately and begin to dominate the runtime, reducing the relative contribution of forward computation. This shift increasingly constrains overall performance under heavy load. 

\begin{figure}[t]
    \centering
    \setlength{\abovecaptionskip}{-1pt}
    \setlength{\belowcaptionskip}{-10pt}
    \includegraphics[width=\columnwidth]{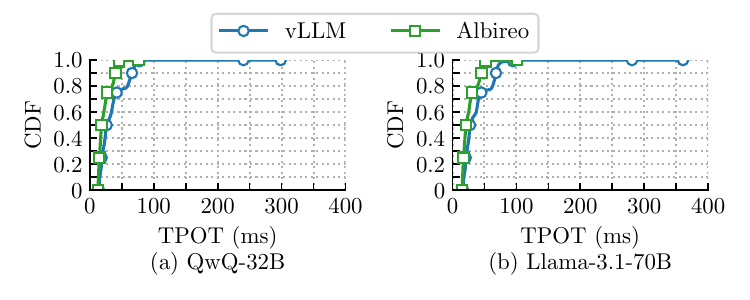}
    \caption{Cumulative distribution of inference latency across models and inference engines, aggregated over one week.}
    \label{fig:production_latency}
\end{figure}

\para{Latency distribution.} In online inference scenarios, tail latency—especially the 99\% and 99.9\% percentiles—is critical for ensuring user experience, in addition to average latency. Figure~\ref{fig:production_latency} presents the TPOT distribution for QwQ-32B and Llama3.1-70B over a one-week period in production, showing that \sysname substantially reduces tail latency.

For example, QwQ-32B's 99\textsuperscript{th} and 99.9\textsuperscript{th} percentile TPOT under vLLM are 271 ms and 298 ms, while \sysname reduces them to 75ms and 76ms. This improvement stems from two key factors: first, \sysname achieves shorter iteration latency through more efficient CPU-GPU parallelism; second, existing inference engines, typically implemented in Python, rely on \emph{asyncio}~\cite{asyncio} to simulate multithreading for CPU tasks. However, the OS scheduling and event-loop overhead cause performance variability in asyncio tasks. Since existing engines do not overlap CPU and GPU tasks, this variability increases end-to-end latency. In contrast, \sysname’s overlapping mechanism effectively hides CPU-side fluctuations, resulting in more stable tail latency.

\begin{figure}
    \centering
    \setlength{\abovecaptionskip}{-1pt}
    \setlength{\belowcaptionskip}{-10pt}
    \includegraphics[width=\columnwidth]{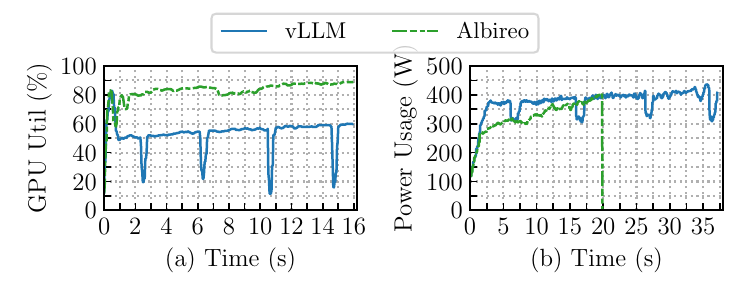}
    \caption{GPU utilization and power usage on Qwen-2.5-32B ($t=4$) on the $H100^{N}$ testbed.}
    \label{fig:line_util_compare}
\end{figure}
\begin{figure}[t]
    \centering
    \setlength{\abovecaptionskip}{-1pt}
    \setlength{\belowcaptionskip}{-10pt}
    \includegraphics[width=\columnwidth]{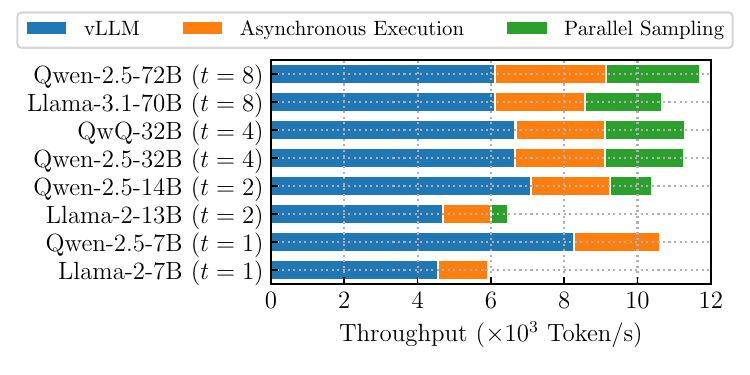}
    \caption{Throughput ablation study of \sysname's optimizations across various models on the $H100^{N}$ testbed.}
    \label{fig:stackh_ablation}
\end{figure}

\subsection{Resource Usage}
\label{sec:resource}

\para{\sysname optimizes GPU utilization of inference systems.}
A key design goal of \sysname is to improve GPU utilization. Our measurements show that by reducing GPU idle time in stages $T_1$, $T_2$, and $T_5$ through asynchronous execution, and by distributing the sampling workload across all available GPUs, \sysname increases average GPU utilization by about 10\% to 40\%. Figure~\ref{fig:line_util_compare}(a) presents the average GPU utilization measured every 20 ms, comparing \sysname with vLLM. As an example, \sysname improves average GPU utilization by 28\% on Qwen-2.5-32B with the default configuration.

\para{\sysname completes inference tasks with less energy.}
While achieving higher GPU utilization, \sysname also enables more energy-efficient inference. Figure~\ref{fig:line_util_compare}(b) shows the average GPU power consumption recorded every 20 ms during inference on 256 concurrent requests using different frameworks. Compared to vLLM, \sysname reduces average GPU power by 15\%, shortens inference time by 47\%, and achieves an overall energy saving of 54\%.

\begin{table}[tb]
    \centering
\setlength{\abovecaptionskip}{5pt}  
\setlength{\belowcaptionskip}{5pt}   
    \caption{Single-task time reduction for Qwen-2.5-32B at $t=4$ on the $H100^{N}$ testbed, achieved by \sysname.}
    \label{tab:breakdown}
    \begin{tabular}{|c|c|c|c|c|}
    \hline 
    Task                   &   $T_{1}$      &   $T_{2}$          &   $T_{4}$         &   $T_{5}$         \\ \hline
    vLLM                   &   $\sim 4\ ms$   &   $\sim 4 \ ms$       &   $\sim 6 \ ms$      &   $\sim 0.5 \ ms$    \\
    \sysname               &   $\sim 5\ \mu s$    &   $\sim 40 \ \mu s$   &   $\sim 1.5 \ ms$    &   $\sim 25 \ \mu s$  \\ \hline
    \end{tabular}
    \crunch
    \crunch
    \crunch
\end{table}

This 15\% power reduction aligns with prior work~\cite{hong2010eneregy}, which shows that GPU energy efficiency (performance per Watt) drops when memory request rates exceed peak bandwidth. Without balancing memory-intensive tasks among GPUs, an inference system may consume more power while delivering lower computational throughput.

\subsection{Ablation Studies}
\label{sec:studies}

\para{Performance impact of the asynchronous execution and sequence-parallel sampling.} 
Figure~\ref{fig:stackh_ablation} illustrates the sources of throughput improvements for each LLM under the default configuration. The bar vLLM represents the throughput of vanilla vLLM. \emph{Asynchronous Execution} and \emph{Parallel Sampling} represent the respective proportions of throughput contributed by the two mechanisms in the overall throughput.

For tiny and small LLMs, the performance gains come almost entirely from \emph{Asynchronous Execution}, as parallel sampling is either disabled or provides limited benefits due to insufficient parallelism. In contrast, for moderate and large LLMs, both mechanisms contribute comparably to the overall performance improvements.

\para{Performance impact of \sysname optimizations on inference tasks.}
\sysname demonstrates consistent performance across all models, significantly reducing non-scalable overhead in each iteration. Table~\ref{tab:breakdown} shows the performance improvements achieved by \sysname on Qwen-2.5-32B at each task of the inference workflow. Since sampling is a hybrid task involving both CPU and GPU computation, we introduce a GPU-CPU synchronization before sampling to isolate the influence of forward computation.

We highlight the following key observations: (1) asynchronous execution in \sysname reduces the time that CPU tasks block the program from 8.5 ms to a negligible 70 $\mu$s, achieving a reduction of over 99\%; (2) With 4 GPUs, parallel sampling saves 75\% of the time in $T_4$. Combined, the two parallel mechanisms reduce the non-scalable overhead in each iteration by more than 89\%.

\label{sec:choice_validation}

\begin{figure}[t]
    \centering
    \setlength{\abovecaptionskip}{-1pt}
    \setlength{\belowcaptionskip}{-5pt}
    \includegraphics[width=0.85\columnwidth]{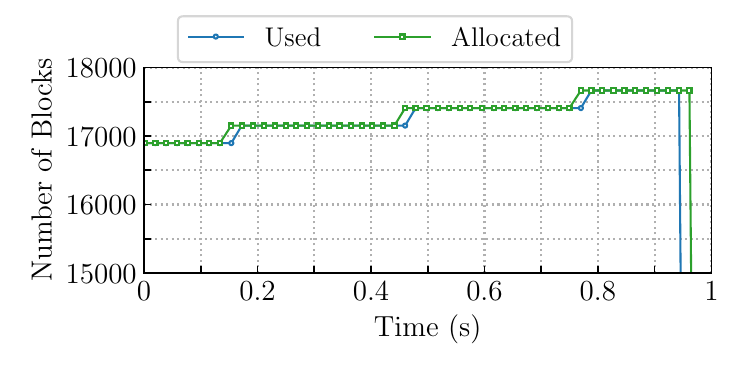}
    \caption{Blocks allocated by \sysname compared to worst-case usage.}
    \label{fig:block_usage}
    \crunch
    \crunch
    \crunch
\end{figure}

\para{Optimistic single-iteration scheduling incurs minimal extra GPU memory usage in the worst case.}
As \sysname's scheduler always assumes sequence continuation, a sequence that stops early can at worst waste one KV block. To assess this, we measure actual GPU memory usage and allocated memory per iteration (Figure~\ref{fig:block_usage} shows the results from the last second for Qwen-2.5-32B on the $H100^{N}$ testbed). Every $B_c$ tokens ($B_c{=}16$ in \sysname and vLLM) trigger preallocation of one block; thereafter, used and allocated blocks match, and a new block is allocated only when the current one fills. If a sequence completes exactly when a block is preallocated, the surplus is reclaimed within one iteration.

\begin{figure}[t]
    \centering
    \setlength{\abovecaptionskip}{-1pt}
    \setlength{\belowcaptionskip}{-10pt}
    \includegraphics[height=4cm, width=0.75\columnwidth, keepaspectratio=false]{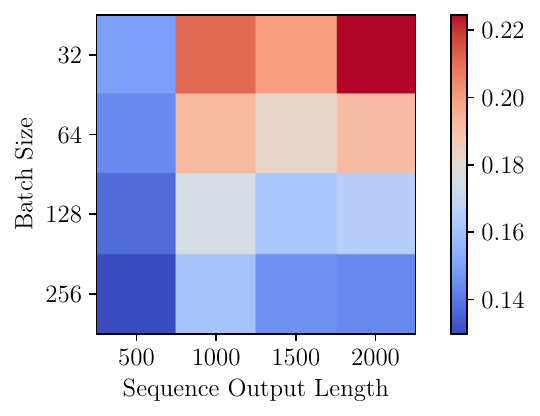}
    \caption{$R_s$ on Qwen-2.5-32B at $t=4$ using $H100^{N}$ testbed.}
    \label{fig:heat}
    \crunch
    \crunch
\end{figure}

\para{Forward computation overlaps extra scattering.}
To validate our claim that the forward pass can effectively hide the overhead of sampling metadata scattering, we measure the $R_s$ (\emph{ratio} of sampling metadata scattering time to forwarding pass time) as the batch size and sequence length increase in Figure~\ref{fig:heat}, where warmer colors indicate higher $R_s$.

Across all configurations on the $H100^{N}$ testbed, we observe that the $R_s$ is significantly small, accounting for only 12\% to 22\%. Even with future devices offering improved performance, we estimate that the ratio of PCIe bandwidth to GPU computational power would need to exceed $4\times$ the current level to challenge this assumption. Thus, our overlapping technique is expected to remain effective in most scenarios.

%% file: 8_conclusion.tex
\section{Limitations and Future Work}
\label{sec:limitation_future}

Looking ahead, we expect that rapidly increasing GPU computational capability—outpacing improvements in CPU technology—will continue to shrink the relative cost of forward passes within each iteration. For example, our evaluation shows that upgrading from A100 to H100 substantially boosts compute throughput while leaving non-scalable components largely unchanged, making these overheads a growing fraction of end-to-end latency. At the same time, steady growth in per-GPU memory capacity across generations is pushing model deployments increasingly toward single-node settings. Together, these trends underscore the rising importance of reducing non-scalable overheads and optimizing intra-node parallel execution as GPU compute becomes less of a bottleneck.

\sysname focuses on efficient inference within a single node, leveraging parallelism and asynchronous execution to reduce these non-scalable overheads. However, its current design does not directly address multi-node deployments for ultra-large models, which introduce fundamentally different scalability challenges. In such environments, TP all-reduce operations often dominate execution time, and hybrid TP–PP brings additional issues such as inter-stage communication bubbles and stage load imbalance (e.g., extra sampling work in the final stage). These challenges are largely orthogonal to the intra-node optimizations explored in this work. Future work will extend \sysname to hybrid TP–PP multi-node deployments, addressing inter-stage communication efficiency and improving load balance across pipeline stages.

%% file: Appendix.tex
\section{Parallel Output Processing}
\label{sec:parallel_output_processing}

\begin{figure}[t]
    \centering
    \includegraphics[width=\columnwidth]{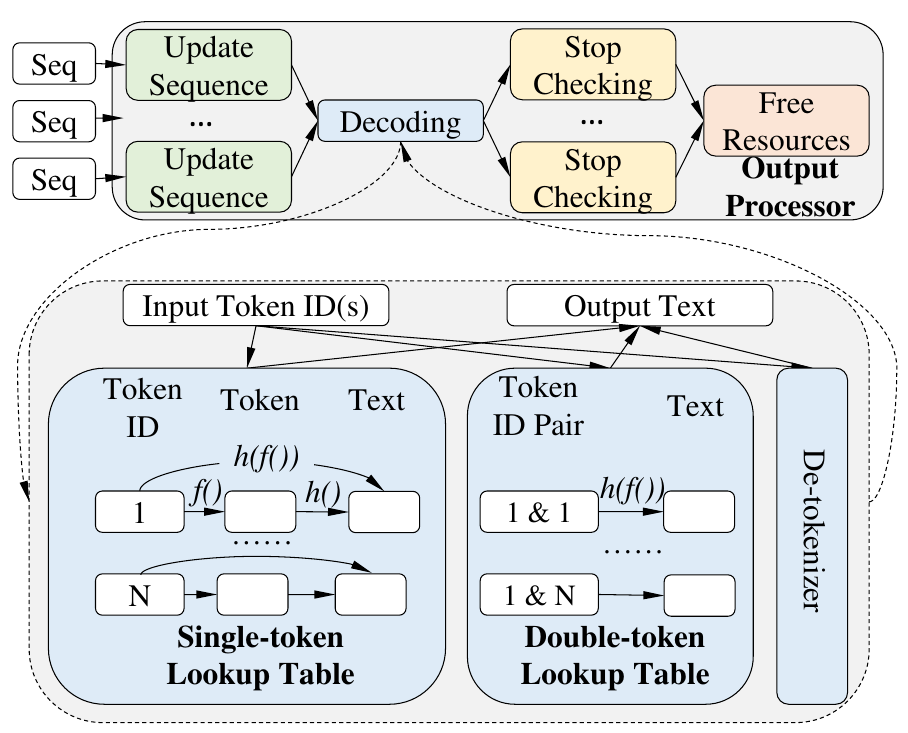}
    \caption{Parallelized output processing.}
    \label{fig:output_proc}
\end{figure}

Since existing inference frameworks~\cite{kwon2023vllm, zheng2024sglang, DeepSpeed} typically implement output processing via a python loop, they process each batch of sequences in a sequential manner. Consequently, the combined duration of scheduling, input processing, and output processing may, in some instances, exceed that of decoding forward, thereby impeding the effective overlap of CPU tasks with GPU operations.

We break down the output processor's sequence processing into four distinct steps and optimize each step individually. As illustrated in Figure~\ref{fig:output_proc}, the output processor handles a sequence through four steps: 
\begin{itemize}
    \item update sequence: The newly sampled token ID and its corresponding probability are appended to the sequence.  
    \item decoding: Each sequence is incrementally decoded using the new sampled token ID.  
    \item stop checking: The inference framework evaluates whether the sequence meets completion conditions, such as generating a specific string, encountering a special token ID, or reaching the maximum length.  
    \item free resources: Resources associated with finished sequences are released, and the generated results are returned.
\end{itemize}
The \emph{update sequence} and \emph{stop checking} steps for different sequences are entirely independent, allowing \sysname to utilize thread pools for parallel execution of these calculations. In contrast, \emph{free resources} depends on the KV cache management module and must be executed serially. However, since each sequence releases resources only once at the end of its lifecycle, the infrequent triggering of \emph{free resources} has a negligible impact on overall performance.

\emph{Decoding} is a crucial and time-consuming step in output processing for each sequence. However, because it depends on the thread-unsafe \emph{de-tokenizer}, it must be executed sequentially. \sysname mitigates the time-consuming decoding process by using two fast \emph{lookup} tables, thereby preventing frequent calls to the \emph{de-tokenizer}.

Let us first briefly explain the function of the \emph{de-tokenizer}: Given a sequence \( s = \langle id_1, id_2, \dots, id_n \rangle \), the \emph{de-tokenizer} performs two decoding functions:
$$
f(\langle id_1, id_2, \dots, id_n \rangle) = \langle token_1, token_2, \dots, token_n \rangle
$$
and
$$
h(\langle token_1, token_2, \dots, token_n \rangle) = text
$$
where
$$
f(\langle id_i, id_j \rangle) = \langle f(id_i), f(id_j) \rangle
$$
and
$$
h(\langle token_i, token_j \rangle) \ne h(token_i) + h(token_j)
$$
To reduce performance overhead, \sysname employs incremental decoding. Specifically, assuming the current output text of sequence \( s \) is \( text_n \), when a new sampled token ID \( id_{n+1} \) arrives, \sysname calculates the incremental text as:
\begin{equation}
\label{equ:decoding}
text \gets h(\langle f(id_n), f(id_{n+1}) \rangle) - h(f(id_n))
\end{equation}
and updates the output text by:
$$
text_{n+1} \gets text_n + text
$$

We derive two key insights from Equation~\ref{equ:decoding}:  
\begin{itemize}
    \item Single-Token Lookup: The computation of \( h(f(id_n)) \) can be reduced to an \( O(1) \) lookup in a table mapping token IDs to their corresponding text. This simplification is feasible because token IDs in LLMs are continuous and finite. For example, Qwen-2.5-14B has 152,064 token IDs ranging from 0 to 152,063, enabling \sysname to cache all tokens in a single-token lookup table (Figure~\ref{fig:output_proc}).  
    \item Double-Token Lookup: While the value space of \( id_n \) and \( id_{n+1} \) pairs in \( h(\langle f(id_n), f(id_{n+1}) \rangle) \) is large and impractical to cache fully, Zipf's law~\cite{piantadosi2014zipf} suggests that many token pairs occur infrequently in natural language. Therefore, \( h(\langle f(id_n), f(id_{n+1}) \rangle) \) can be cached by storing commonly used pairs. In \sysname, we cache \( 10^9 \) pairs (take about 30 GB memory) for each model (double-token lookup table in Figure~\ref{fig:output_proc}), achieving a hit rate of over 99\%. When a lookup fails, the \emph{de-tokenizer} is invoked to perform the decoding operation.
\end{itemize}

These optimizations enable the output processing task to complete within 2 ms per iteration, ensuring seamless overlap between GPU and CPU tasks.






%% file: ref.bib
@misc{asyncio,
    title = {{Asynchronous I/O}},
    key = {asyncio},
    howpublished = {\url{https://docs.python.org/3/library/asyncio.html}},
    author = {Python},
    year = {2025}
}

@misc{vllm_async,
    title = {{vLLM Asynchronous Scheduling}},
    key = {vllmasync},
    howpublished = {\url{https://github.com/vllm-project/vllm/pull/24799}},
    author = {vLLM},
    year = {2025}
}

@misc{sglang_zero,
    title = {SGLang: Zero-Overhead Batch Scheduler},
    key = {zero},
    howpublished = {\url{https://lmsys.org/blog/2024-12-04-sglang-v0-4/}},
    author = {SGLang},
    year = {2024}
}

@inproceedings{zhu2025nanoflow,
  title={$\{$NanoFlow$\}$: Towards optimal large language model serving throughput},
  author={Zhu, Kan and Gao, Yufei and Zhao, Yilong and Zhao, Liangyu and Zuo, Gefei and Gu, Yile and Xie, Dedong and Ye, Zihao and Kamahori, Keisuke and Lin, Chien-Yu and others},
  booktitle={19th USENIX Symposium on Operating Systems Design and Implementation (OSDI 25)},
  pages={749--765},
  year={2025}
}

@article{shoeybi2019megatron,
  title={Megatron-lm: Training multi-billion parameter language models using model parallelism},
  author={Shoeybi, Mohammad and Patwary, Mostofa and Puri, Raul and LeGresley, Patrick and Casper, Jared and Catanzaro, Bryan},
  journal={arXiv preprint arXiv:1909.08053},
  year={2019}
}

@article{piantadosi2014zipf,
  title={Zipf’s word frequency law in natural language: A critical review and future directions},
  author={Piantadosi, Steven T},
  journal={Psychonomic bulletin \& review},
  volume={21},
  number={5},
  pages={1112--1130},
  year={2014},
  publisher={Springer}
}

@misc{llama-2-7b,
    title = {{meta-llama/Llama-2-7b}},
    key = {metallama},
    howpublished = {\url{https://huggingface.co/meta-llama/Llama-2-7b}},
    author = {Meta},
    year = {2025}
}

@misc{llama-2-13b-hf,
      title={Llama 2: Open Foundation and Fine-Tuned Chat Models}, 
      author={Hugo Touvron and Louis Martin and Kevin Stone and Peter Albert and Amjad Almahairi and Yasmine Babaei and Nikolay Bashlykov and Soumya Batra and Prajjwal Bhargava and Shruti Bhosale and Dan Bikel and Lukas Blecher and Cristian Canton Ferrer and Moya Chen and Guillem Cucurull and David Esiobu and Jude Fernandes and Jeremy Fu and Wenyin Fu and Brian Fuller and Cynthia Gao and Vedanuj Goswami and Naman Goyal and Anthony Hartshorn and Saghar Hosseini and Rui Hou and Hakan Inan and Marcin Kardas and Viktor Kerkez and Madian Khabsa and Isabel Kloumann and Artem Korenev and Punit Singh Koura and Marie-Anne Lachaux and Thibaut Lavril and Jenya Lee and Diana Liskovich and Yinghai Lu and Yuning Mao and Xavier Martinet and Todor Mihaylov and Pushkar Mishra and Igor Molybog and Yixin Nie and Andrew Poulton and Jeremy Reizenstein and Rashi Rungta and Kalyan Saladi and Alan Schelten and Ruan Silva and Eric Michael Smith and Ranjan Subramanian and Xiaoqing Ellen Tan and Binh Tang and Ross Taylor and Adina Williams and Jian Xiang Kuan and Puxin Xu and Zheng Yan and Iliyan Zarov and Yuchen Zhang and Angela Fan and Melanie Kambadur and Sharan Narang and Aurelien Rodriguez and Robert Stojnic and Sergey Edunov and Thomas Scialom},
      year={2023},
      eprint={2307.09288},
      archivePrefix={arXiv},
      primaryClass={cs.CL},
      url={https://arxiv.org/abs/2307.09288}, 
}

@misc{qwen2.5,
    title = {Qwen2.5: A Party of Foundation Models},
    url = {https://qwenlm.github.io/blog/qwen2.5/},
    author = {Qwen Team},
    month = {September},
    year = {2024}
}

@misc{qwq,
    title = {{QwQ: Reflect deeply on the boundaries of the unknown}},
    key = {qwq},
    howpublished = {\url{https://qwenlm.github.io/blog/qwq-32b-preview/}},
    author = {Qwen Team},
    year = {2025}
}

@misc{llama-3.1,
      title={The Carbon Footprint of Machine Learning Training Will Plateau, Then Shrink}, 
      author={David Patterson and Joseph Gonzalez and Urs Hölzle and Quoc Le and Chen Liang and Lluis-Miquel Munguia and Daniel Rothchild and David So and Maud Texier and Jeff Dean},
      year={2022},
      eprint={2204.05149},
      archivePrefix={arXiv},
      primaryClass={cs.LG},
      url={https://arxiv.org/abs/2204.05149}, 
}

@inproceedings{prabhu2025vattention,
  title={vattention: Dynamic memory management for serving llms without pagedattention},
  author={Prabhu, Ramya and Nayak, Ajay and Mohan, Jayashree and Ramjee, Ramachandran and Panwar, Ashish},
  booktitle={Proceedings of the 30th ACM International Conference on Architectural Support for Programming Languages and Operating Systems, Volume 1},
  pages={1133--1150},
  year={2025}
}

@inproceedings{liu2024cachegen,
  title={Cachegen: Kv cache compression and streaming for fast large language model serving},
  author={Liu, Yuhan and Li, Hanchen and Cheng, Yihua and Ray, Siddhant and Huang, Yuyang and Zhang, Qizheng and Du, Kuntai and Yao, Jiayi and Lu, Shan and Ananthanarayanan, Ganesh and others},
  booktitle={Proceedings of the ACM SIGCOMM 2024 Conference},
  pages={38--56},
  year={2024}
}

@inproceedings{chitty2024llm,
  title={LLM-Inference-Bench: Inference Benchmarking of Large Language Models on AI Accelerators},
  author={Chitty-Venkata, Krishna Teja and Raskar, Siddhisanket and Kale, Bharat and Ferdaus, Farah and Tanikanti, Aditya and Raffenetti, Ken and Taylor, Valerie and Emani, Murali and Vishwanath, Venkatram},
  booktitle={SC24-W: Workshops of the International Conference for High Performance Computing, Networking, Storage and Analysis},
  pages={1362--1379},
  year={2024},
  organization={IEEE}
}

@article{vaswani2017transformer,
  title={Attention is all you need},
  author={Vaswani, Ashish and Shazeer, Noam and Parmar, Niki and Uszkoreit, Jakob and Jones, Llion and Gomez, Aidan N and Kaiser, {\L}ukasz and Polosukhin, Illia},
  journal={Advances in neural information processing systems},
  volume={30},
  year={2017}
}

@misc{DeepSpeed,
  key       = {DeepSpeed},
  title     = {DeepSpeed},
  note      = {\url{https://github.com/microsoft/DeepSpeed}},
  year      = 2024,
  author    = {Microsoft},
}

@article{korthikanti2023megatron2,
  title={Reducing activation recomputation in large transformer models},
  author={Korthikanti, Vijay Anand and Casper, Jared and Lym, Sangkug and McAfee, Lawrence and Andersch, Michael and Shoeybi, Mohammad and Catanzaro, Bryan},
  journal={Proceedings of Machine Learning and Systems},
  volume={5},
  year={2023}
}

@article{ouyang2022gpt,
  title={Training language models to follow instructions with human feedback},
  author={Ouyang, Long and Wu, Jeffrey and Jiang, Xu and Almeida, Diogo and Wainwright, Carroll and Mishkin, Pamela and Zhang, Chong and Agarwal, Sandhini and Slama, Katarina and Ray, Alex and others},
  journal={Advances in Neural Information Processing Systems},
  volume={35},
  pages={27730--27744},
  year={2022}
}

@misc{pickle,
  key       = {pickle},
  title     = {pickle — Python object serialization},
  note      = {\url{https://docs.python.org/3/library/pickle.html}},
  year      = 2024,
  author    = {Python},
}

@misc{vllm_llama3,
    key = {vllm_llama3},
    title = {vLLM: Optimization and Tuning},
    note = {\url{https://docs.vllm.ai/en/latest/configuration/optimization.html}},
    year = {2025},
    author = {vLLM}
}

@article{bhaskar2025cache,
  title={Cache Me If You Can: How Many KVs Do You Need for Effective Long-Context LMs?},
  author={Bhaskar, Adithya and Wettig, Alexander and Gao, Tianyu and Dong, Yihe and Chen, Danqi},
  journal={arXiv preprint arXiv:2506.17121},
  year={2025}
}

@article{penalty-rep,
  title={Importance of a search strategy in neural dialogue modelling},
  author={Kulikov, Ilya and Miller, Alexander H and Cho, Kyunghyun and Weston, Jason},
  journal={arXiv preprint arXiv:1811.00907},
  volume={2},
  year={2018},
  publisher={Nov}
}

@misc{penalty-fre-pre,
    title = {{OpenAI API Documentation}},
    key = {penaltyfrepre},
    howpublished = {https://platform.openai.com/docs},
    author = {OpenAI},
    year = {2025}
}

@misc{vllm_tp,
    key = {vllmtp},
    title = {vLLM: Parallelism and Scaling},
    note = {\url{https://docs.vllm.ai/en/latest/serving/parallelism_scaling.html}},
    year = {2025},
    author = {vLLM}
}

@misc{sglang_doc,
    key = {sglangdoc},
    title = {SGLang Documentation},
    note = {\url{https://docs.sglang.ai/index.html}},
    year = {2025},
    author = {SGLang}
}

@inproceedings{kwon2023vllm,
  title={Efficient memory management for large language model serving with pagedattention},
  author={Kwon, Woosuk and Li, Zhuohan and Zhuang, Siyuan and Sheng, Ying and Zheng, Lianmin and Yu, Cody Hao and Gonzalez, Joseph and Zhang, Hao and Stoica, Ion},
  booktitle={Proceedings of the 29th Symposium on Operating Systems Principles},
  pages={611--626},
  year={2023}
}

@inproceedings{yu2022orca,
  title={Orca: A distributed serving system for  Transformer-Based  generative models},
  author={Yu, Gyeong-In and Jeong, Joo Seong and Kim, Geon-Woo and Kim, Soojeong and Chun, Byung-Gon},
  booktitle={16th USENIX Symposium on Operating Systems Design and Implementation (OSDI 22)},
  pages={521--538},
  year={2022}
}

@misc{sglang_bench,
    author = {SGLang},
    key = {sglangbench},
    title = {SGLang: Benchmark and Profiling},
    howpublished = {\url{https://docs.sglang.ai/developer_guide/benchmark_and_profiling.html}},
    year = {2025}
}

@inproceedings {zhong2024distserve,
author = {Yinmin Zhong and Shengyu Liu and Junda Chen and Jianbo Hu and Yibo Zhu and Xuanzhe Liu and Xin Jin and Hao Zhang},
title = {DistServe: Disaggregating Prefill and Decoding for Goodput-optimized Large Language Model Serving},
booktitle = {18th USENIX Symposium on Operating Systems Design and Implementation (OSDI 24)},
year = {2024},
isbn = {978-1-939133-40-3},
address = {Santa Clara, CA},
pages = {193--210},
url = {https://www.usenix.org/conference/osdi24/presentation/zhong-yinmin},
publisher = {USENIX Association},
month = jul
}

@inproceedings{gao2024cost,
  title={ Cost-Efficient  Large Language Model Serving for Multi-turn Conversations with  CachedAttention },
  author={Gao, Bin and He, Zhuomin and Sharma, Puru and Kang, Qingxuan and Jevdjic, Djordje and Deng, Junbo and Yang, Xingkun and Yu, Zhou and Zuo, Pengfei},
  booktitle={2024 USENIX Annual Technical Conference (USENIX ATC 24)},
  pages={111--126},
  year={2024}
}

@article{cai2024pyramidkv,
  title={Pyramidkv: Dynamic kv cache compression based on pyramidal information funneling},
  author={Cai, Zefan and Zhang, Yichi and Gao, Bofei and Liu, Yuliang and Liu, Tianyu and Lu, Keming and Xiong, Wayne and Dong, Yue and Chang, Baobao and Hu, Junjie and others},
  journal={arXiv preprint arXiv:2406.02069},
  year={2024}
}

@article{xiao2023sink,
  title={Efficient streaming language models with attention sinks},
  author={Xiao, Guangxuan and Tian, Yuandong and Chen, Beidi and Han, Song and Lewis, Mike},
  journal={arXiv preprint arXiv:2309.17453},
  year={2023}
}

@article{li2024snapkv,
  title={Snapkv: {LLM} knows what you are looking for before generation},
  author={Li, Yuhong and Huang, Yingbing and Yang, Bowen and Venkitesh, Bharat and Locatelli, Acyr and Ye, Hanchen and Cai, Tianle and Lewis, Patrick and Chen, Deming},
  journal={arXiv preprint arXiv:2404.14469},
  year={2024}
}

@article{zheng2024sglang,
  title={Sglang: Efficient execution of structured language model programs},
  author={Zheng, Lianmin and Yin, Liangsheng and Xie, Zhiqiang and Sun, Chuyue and Huang, Jeff and Yu, Cody Hao and Cao, Shiyi and Kozyrakis, Christos and Stoica, Ion and Gonzalez, Joseph E and others},
  journal={arXiv preprint arXiv:2312.07104},
  year={2024}
}

@inproceedings{fu2024serverlessllm,
  title={ServerlessLLM: Low-latency serverless inference for large language models},
  author={Fu, Yao and Xue, Leyang and Huang, Yeqi and Brabete, Andrei-Octavian and Ustiugov, Dmitrii and Patel, Yuvraj and Mai, Luo},
  booktitle={18th USENIX Symposium on Operating Systems Design and Implementation (OSDI 24)},
  pages={135--153},
  year={2024},
  organization={USENIX Association}
}

@inproceedings {sun2024llumnix,
author = {Biao Sun and Ziming Huang and Hanyu Zhao and Wencong Xiao and Xinyi Zhang and Yong Li and Wei Lin},
title = {Llumnix: Dynamic Scheduling for Large Language Model Serving},
booktitle = {18th USENIX Symposium on Operating Systems Design and Implementation (OSDI 24)},
year = {2024},
isbn = {978-1-939133-40-3},
address = {Santa Clara, CA},
pages = {173--191},
url = {https://www.usenix.org/conference/osdi24/presentation/sun-biao},
publisher = {USENIX Association},
month = jul
}

@article{gage1994new,
  title={A new algorithm for data compression},
  author={Gage, Philip},
  journal={The C Users Journal},
  volume={12},
  number={2},
  pages={23--38},
  year={1994},
  publisher={R \& D Publications, Inc. Lawrence, KS, USA}
}

@inproceedings{kudo-richardson-2018-sentencepiece,
    title = "{S}entence{P}iece: A simple and language independent subword tokenizer and detokenizer for Neural Text Processing",
    author = "Kudo, Taku  and
      Richardson, John",
    editor = "Blanco, Eduardo  and
      Lu, Wei",
    booktitle = "Proceedings of the 2018 Conference on Empirical Methods in Natural Language Processing: System Demonstrations",
    month = nov,
    year = "2018",
    address = "Brussels, Belgium",
    publisher = "Association for Computational Linguistics",
    url = "https://aclanthology.org/D18-2012",
    doi = "10.18653/v1/D18-2012",
    pages = "66--71",
}

@inproceedings{sennrich-etal-2016-neural,
    title = "Neural Machine Translation of Rare Words with Subword Units",
    author = "Sennrich, Rico  and
      Haddow, Barry  and
      Birch, Alexandra",
    editor = "Erk, Katrin  and
      Smith, Noah A.",
    booktitle = "Proceedings of the 54th Annual Meeting of the Association for Computational Linguistics (Volume 1: Long Papers)",
    month = aug,
    year = "2016",
    address = "Berlin, Germany",
    publisher = "Association for Computational Linguistics",
    url = "https://aclanthology.org/P16-1162",
    doi = "10.18653/v1/P16-1162",
    pages = "1715--1725",
}

@misc{databricks,
  key       = {Databricks-dolly-15k},
  title     = {Databricks-dolly-15k is an open source dataset of instruction-following records generated by thousands of {Databricks} employees in several of the behavioral categories outlined in the {InstructGPT} paper},
  note      = {\url{https://huggingface.co/datasets/databricks/databricks-dolly-15k}},
  year      = 2024,
  author    = {Databricks},
}

@misc{bentoML,
  key       = {bentoML},
  title     = {Benchmarking {LLM} Inference Backends: {vLLM}, {LMDeploy}, {MLC-LLM}, {TensorRT-LLM}, and {TGI}},
  note      = {\url{https://bentoml.com/blog/benchmarking-llm-inference-backends}},
  year      = 2024,
  author    = {Rick Zhou},
}

@inproceedings{hong2010eneregy,
  title={An integrated {GPU} power and performance model},
  author={Hong, Sunpyo and Kim, Hyesoon},
  booktitle={Proceedings of the 37th annual international symposium on Computer architecture},
  pages={280--289},
  year={2010}
}

@article{team2025kimik2,
  title={Kimi K2: Open Agentic Intelligence},
  author={Team, Kimi and Bai, Yifan and Bao, Yiping and Chen, Guanduo and Chen, Jiahao and Chen, Ningxin and Chen, Ruijue and Chen, Yanru and Chen, Yuankun and Chen, Yutian and others},
  journal={arXiv preprint arXiv:2507.20534},
  year={2025}
}

@inproceedings{qin2025mooncake,
  title={Mooncake: Trading more storage for less computation—a $\{$KVCache-centric$\}$ architecture for serving $\{$LLM$\}$ chatbot},
  author={Qin, Ruoyu and Li, Zheming and He, Weiran and Cui, Jialei and Ren, Feng and Zhang, Mingxing and Wu, Yongwei and Zheng, Weimin and Xu, Xinran},
  booktitle={23rd USENIX Conference on File and Storage Technologies (FAST 25)},
  pages={155--170},
  year={2025}
}

@book{amdahl1967Amdahl,
  title={Computer architecture: a quantitative approach},
  author={Hennessy, John L and Patterson, David A},
  year={2011},
  publisher={Elsevier}
}

@misc{llama2deploy,
    title={{Llama 2}},
    key = {llama2deploy},
    howpublished = {\url{https://infohub.delltechnologies.com/ja-jp/l/llama-2-inferencing-on-a-single-gpu/introduction-3976/}},
    author = {Dell},
    year = {2025}
}

@article{zhang2023h2o,
  title={H2o: Heavy-hitter oracle for efficient generative inference of large language models},
  author={Zhang, Zhenyu and Sheng, Ying and Zhou, Tianyi and Chen, Tianlong and Zheng, Lianmin and Cai, Ruisi and Song, Zhao and Tian, Yuandong and R{\'e}, Christopher and Barrett, Clark and others},
  journal={Advances in Neural Information Processing Systems},
  volume={36},
  pages={34661--34710},
  year={2023}
}

@article{liu2024deepseek,
  title={Deepseek-v2: A strong, economical, and efficient mixture-of-experts language model},
  author={Liu, Aixin and Feng, Bei and Wang, Bin and Wang, Bingxuan and Liu, Bo and Zhao, Chenggang and Dengr, Chengqi and Ruan, Chong and Dai, Damai and Guo, Daya and others},
  journal={arXiv preprint arXiv:2405.04434},
  year={2024}
}

@article{liu2024deepseekv3,
  title={Deepseek-v3 technical report},
  author={Liu, Aixin and Feng, Bei and Xue, Bing and Wang, Bingxuan and Wu, Bochao and Lu, Chengda and Zhao, Chenggang and Deng, Chengqi and Zhang, Chenyu and Ruan, Chong and others},
  journal={arXiv preprint arXiv:2412.19437},
  year={2024}
}

@inproceedings {2024userve,
author = {Haoran Qiu and Weichao Mao and Archit Patke and Shengkun Cui and Saurabh Jha and Chen Wang and Hubertus Franke and Zbigniew Kalbarczyk and Tamer Ba{\c s}ar and Ravishankar K. Iyer},
title = {Power-aware Deep Learning Model Serving with {u-Serve}},
booktitle = {2024 USENIX Annual Technical Conference (USENIX ATC 24)},
year = {2024},
isbn = {978-1-939133-41-0},
address = {Santa Clara, CA},
pages = {75--93},
url = {https://www.usenix.org/conference/atc24/presentation/qiu},
publisher = {USENIX Association},
month = jul
}

@inproceedings {2024Sarathi,
author = {Amey Agrawal and Nitin Kedia and Ashish Panwar and Jayashree Mohan and Nipun Kwatra and Bhargav Gulavani and Alexey Tumanov and Ramachandran Ramjee},
title = {Taming Throughput-Latency Tradeoff in LLM Inference with Sarathi-Serve},
booktitle = {18th USENIX Symposium on Operating Systems Design and Implementation (OSDI 24)},
year = {2024},
isbn = {978-1-939133-40-3},
address = {Santa Clara, CA},
pages = {117--134},
url = {https://www.usenix.org/conference/osdi24/presentation/agrawal},
publisher = {USENIX Association},
month = jul
}

@inproceedings {2024InfiniGen,
author = {Wonbeom Lee and Jungi Lee and Junghwan Seo and Jaewoong Sim},
title = {InfiniGen: Efficient Generative Inference of Large Language Models with Dynamic {KV} Cache Management},
booktitle = {18th USENIX Symposium on Operating Systems Design and Implementation (OSDI 24)},
year = {2024},
isbn = {978-1-939133-40-3},
address = {Santa Clara, CA},
pages = {155--172},
url = {https://www.usenix.org/conference/osdi24/presentation/lee},
publisher = {USENIX Association},
month = jul
}

@article{Ye2025FlashInfer,
  title   = {FlashInfer: Efficient and Customizable Attention Engine for LLM Inference Serving},
  author  = {Zihao Ye and Lequn Chen and Ruihang Lai and Wuwei Lin and Yineng Zhang and Stephanie Wang and Tianqi Chen and Baris Kasikci and Vinod Grover and Arvind Krishnamurthy and Luis Ceze},
  journal = {arXiv preprint arXiv:2501.01005},
  year    = {2025},
  doi     = {10.48550/arXiv.2501.01005},
  url     = {https://arxiv.org/abs/2501.01005}
}

@inproceedings{Kamath2025PODAttention,
  title     = {POD-Attention: Unlocking Full Prefill-Decode Overlap for Faster LLM Inference},
  author    = {Aditya K. Kamath and Ramya Prabhu and Jayashree Mohan and Simon Peter and Ramachandran Ramjee and Ashish Panwar},
  booktitle = {Proceedings of ASPLOS '25},
  address   = {Rotterdam, Netherlands},
  publisher = {ACM},
  year      = {2025},
  url       = {https://www.microsoft.com/en-us/research/wp-content/uploads/2025/03/POD-Attention-ASPLOS25.pdf}
}

@article{Xu2025SpecEE,
  title   = {SpecEE: Accelerating Large Language Model Inference with Speculative Early Exiting},
  author  = {Jiaming Xu and Jiayi Pan and Yongkang Zhou and Siming Chen and Jinhao Li and Yaoxiu Lian and Junyi Wu and Guohao Dai},
  journal = {arXiv preprint arXiv:2504.08850},
  note    = {Accepted by ISCA 2025},
  year    = {2025},
  doi     = {10.48550/arXiv.2504.08850},
  url     = {https://arxiv.org/abs/2504.08850}
}

@article{EntezariZarch2025DEL,
  title   = {DEL: Context-Aware Dynamic Exit Layer for Efficient Self-Speculative Decoding},
  author  = {Hossein Entezari Zarch and Lei Gao and Chaoyi Jiang and Murali Annavaram},
  journal = {arXiv preprint arXiv:2504.05598},
  year    = {2025},
  doi     = {10.48550/arXiv.2504.05598},
  url     = {https://arxiv.org/abs/2504.05598}
}

@article{Yuzuguler2025PRESERVE,
  title   = {PRESERVE: Prefetching Model Weights and KV-Cache in Distributed LLM Serving},
  author  = {Ahmet Caner Y{\"u}z{\"u}g{\"u}ler and Jiawei Zhuang and Lukas Cavigelli},
  journal = {arXiv preprint arXiv:2501.08192},
  year    = {2025},
  doi     = {10.48550/arXiv.2501.08192},
  url     = {https://arxiv.org/abs/2501.08192}
}

@article{Jo2025FastKV,
  title   = {FastKV: KV Cache Compression for Fast Long-Context Processing with Token-Selective Propagation},
  author  = {Dongwon Jo and Jiwon Song and Yulhwa Kim and Jae-Joon Kim},
  journal = {arXiv preprint arXiv:2502.01068},
  year    = {2025},
  doi     = {10.48550/arXiv.2502.01068},
  url     = {https://arxiv.org/abs/2502.01068}
}
